\documentclass[aps,prd,onecolumn,eqsecnum,amssymb,amsmath,showpacs,a4paper, superscriptaddress]{revtex4-2}

\selectfont

\usepackage{graphicx}       
\usepackage{dcolumn}        
\usepackage{bm}             
\usepackage{psfrag}
\usepackage{tensor}
\usepackage[usenames]{color}
\definecolor{navyblue}{rgb}{0.0, 0.0, 0.5}
\usepackage[linktocpage,colorlinks=true,allcolors=navyblue]{hyperref}
\usepackage{amsmath}
\usepackage{caption}
\usepackage{subcaption}

\newcommand{\be}{\begin{equation}}
\newcommand{\ee}{\end{equation}}

\newcommand{\beq}{\begin{equation}}
\newcommand{\eeq}{\end{equation}}

\newcommand{\Equiv}[1]{\underset{#1}{\sim}}
\renewcommand{\vec}[1]{\bm{#1}}
\newcommand{\id}{\bm{1}}

\begin{document}

\preprint{}

\title{Canonical scattering problem in topological metamaterials: 
\\Valley-Hall modes through a bend}


\author{Theo Torres}
\email{torres@lma.cnrs-mrs.fr}
\affiliation{Aix Marseille Univ., CNRS, Centrale Marseille, LMA UMR 7031, Marseille, France}
\author{Cédric Bellis}
\email{bellis@lma.cnrs-mrs.fr}
\affiliation{Aix Marseille Univ., CNRS, Centrale Marseille, LMA UMR 7031, Marseille, France}
\author{Régis Cottereau}
\email{cottereau@lma.cnrs-mrs.fr}
\affiliation{Aix Marseille Univ., CNRS, Centrale Marseille, LMA UMR 7031, Marseille, France}
\author{Antonin Coutant}
\email{coutant@lma.cnrs-mrs.fr}
\affiliation{Aix Marseille Univ., CNRS, Centrale Marseille, LMA UMR 7031, Marseille, France}

\begin{abstract}
We study the amount of backscattering of Valley Hall modes in a classical topological insulator. In reciprocal systems, the conservation of the valley index has been argued to be at the root of the high-transmission of Valley Hall modes, observed in many experimental realisations.
Here, we reconsider this hypothesis by quantitatively analysing the canonical scattering problem of interface Valley Hall modes impinging on sharp bends which may or may not conserve the valley index. 
We consider a tight binding model of graphene ribbons with an interface and compute the reflection and transmission coefficients using a transfer matrix formalism.
We find that, in all configurations considered, the transmission of Valley Hall modes is close to being maximal, even in cases where the valley index is not conserved. Hence there appears to be no correlation between valley conservation and good transmission.
Our results serve as a reference case for the design of Valley Hall type metamaterial.
\end{abstract}

\maketitle

\section{Introduction}

Topological insulators are a class of material, oftentimes constructed from an underlying periodic structure, which exhibit unique wave transport properties. First discovered in the context of electrons in solid states physics, it was then extended to the realm of classical wave: electromagnetic, elastic, acoustic and more~\cite{khanikaev2013photonic,lu2014topological,ozawa2019topological,ma2019topological,zangeneh2020topological,miniaci2021design,yves2022topological}. 
In these systems, a non-trivial topological phase manifests itself by displaying edge waves in frequencies inside the gap of the material. A very appealing property of these edge waves is that they appear to be able to propagate across defects or disorder without undergoing backscattering. 
\bigskip 

Historically, the first example of a topological insulator was the quantum Hall effect~\cite{stone1992quantum}. In this system, a gap can host unidirectional edge waves, with exceptional robustness properties against defects. To obtain the necessary topology, the system must break time reversal invariance, or equivalently, break reciprocity. 
In condensed matter, this is ensured by an external magnetic field. Unfortunately, for classical waves, breaking reciprocity is a hard task, and always requires active materials (i.e. energy input). 
While the Quantum Hall effect is presumably the most studied topological insulator, another type has attracted a lot of attention since its original discovery: the Quantum Spin Hall (QSH) effect~\cite{maciejko2011quantum}. 
The main difference is that the latter can be realized in reciprocal systems. In this case, edge waves always come in pairs moving in opposite direction. 
This time, backscattering immunity is achieved through the spin of the electron, which has the exotic property of inducing a phase shift of $\pi$ when time reversal operation is applied twice. 
This peculiarity is what can guarantee a perfect decoupling between oppositely moving modes~\cite{kane2005quantum}. 
Unfortunately again, classical waves do not naturally possess the equivalent of the electron spin. Hence, many works in metamaterials have been devoted to realizing structures able to mimic that spin with the hope that they would protect their associated topological edge waves from backscattering.
\bigskip 

To obtain the equivalent of an electron spin, two main approaches have been followed. The first one relies on symmetries to obtain degenerate energy levels, acting as pseudo-spins~\cite{wu2015scheme,he2016acoustic,liu2017pseudospins}. This historical analogue QSH effect has been extensively studied theoretically and experimentally in mechanical systems~\cite{PhysRevX.8.031074}, acoustics~\cite{mei2016pseudo,he2016acoustic} and photonics~\cite{yves2017crystalline,yang2018visualization} and it has been established that QSH modes exhibit a reduced backscattering against various type of defects. The second one on the contrary, starts with a configuration displaying several inequivalent Dirac points and break inversion symmetry~\cite{lu2017observation,pal2017edge,lu2018valley,zheng2020acoustic}. That second case, which is the focus of this work, is called the Valley Hall effect. The reason is that the symmetry breaking turns the Dirac points into small gaps with valley shaped dispersion relations. Each valley carries a topological charge, and is essentially decoupled from the others. A typical configuration to illustrate this decoupling is the scattering of valley edge modes on sharp bends and various possible combinations, such as $Z$ and $\Omega$ shaped paths. Similarly to the case of QSH devices, many works have observed a very high transmission of valley Hall modes across these bends, which seems to originate from valley conservation\cite{vila2017observation,liu2018tunable,zhang2018achieving,tian2020dispersion,zhang2019programmable,zhang2022gigahertz,haslinger2022localized}.
While both the analogue QSH and VH effects aim at creating pseudo-spin to protect the topological modes, they do so in fundamentally different ways. The QSH designs are based on a 6-fold rotational symmetry while VH systems break inversion symmetry. The two approaches have been compared in a few specific settings, and their robustness to backscattering seems to depend substantially on the type of defects and their implementations~\cite{orazbayev2019quantitative,deng2019comparison}. This observation indicates that the scattering properties are not solely dependent on the topological properties of the modes but also on the symmetries of the underlying systems.
\bigskip

Along these lines and in the last years, it was established in a couple of works that backscattering immunity is never exact in these time reversal invariant topological insulators with classical waves~\cite{orazbayev2019quantitative,deng2019comparison,arora2021direct,leveque2023quantifying,rosiek2023observation,rechtsman2023reciprocal}. For instance, in the Valley Hall effect decoupling between valleys is always approximate, while, in the QSH effect, the introduction of a defect necessarily breaks the spatial symmetries needed for the construction of the QSH topological invariant. Hence, the question was raised whether the observed high transmission across bends is due to a topological protection or to an \emph{accidental} (i.e. not protected by symmetry) good mode matching on each side of the bend. Such question is particularly relevant, since there exist scattering configurations that inevitably require the non-conservation of the valley index, as we will see below. To settle this question, quantitative analysis of the scattering across bands must be conducted. In this work, we analyse this scattering problem in the context of a tight binding model of a Valley Hall system. Such systems can be seen as a proxy for a photonic or phononic metamaterial but are also pertinent for condensed matter systems.
\bigskip

The motivation of this work is to see this specific model as a canonical scattering problem, which could serve as a reference case for Valley Hall type metamaterials\cite{nethercote2020analytical} and to further understand the necessary conditions for high-transmission across sharp bends~\cite{tang2020observations,makwana2018designing,proctor2020manipulating}.
Here we investigate the propagation of Valley Hall states in graphene ribbons with two possible interfaces. Using a transfer matrix formalism, we quantify the amount of backscattering when Valley Hall states scatter through sharp turns with an angle of $\pi/3$ and $2\pi/3$. The main difference between the two configurations is that it is possible to conserve the valley index for the $2\pi/3$ turn while it is not for the $\pi/3$ case. 
We show that Valley Hall states are well transmitted through the two possible turns, independently of the angle.
Our results support the idea that valley conservation is not correlated to a high transmission across turns.
\bigskip

The paper is organised as follows: In Sec.~\ref{sec:lattice} we discuss the valley Hall effect and its realisation in a lattice model. We describe the band structure and the properties of interface modes in graphene ribbons with an interface. In Sec.~\ref{sec:transfer}, we review the transfer matrix formalism which allows us to quantify the amount of backscattering. 
In Sec.~\ref{sec:analysis} we present the quantitative analysis of the scattering of valley Hall states through a turn at different angles.
We finish our study by a discussion of our results in Sec.~\ref{sec:conclusion}.
The main text is complemented by supplementary materials which contain information about technical details and an extension of our formalism to quantitatively analyse other lattice ribbons with an application to graphene ribbons with armchair edges.

\section{Lattice Model}\label{sec:lattice}

In the following, we will investigate the realisation and scattering of valley Hall modes in a discrete honeycomb lattice system. In this model, the system is described by a set of complex amplitudes associated with each node of the lattice and coupled to their nearest neighbours.
Such approach is commonly used in condensed matter as it describes the tight-binding approximation of the propagation of electrons in crystals.
Discrete models play an important role in modelling continuous classical system in asymptotic regimes such as low-frequency phononics and photonics crystals~\cite{vanel2017asymptotic}, mechanical lattices~\cite{chen2018study} or in the mono-mode approximation of equal-lengths acoustic networks~\cite{coutant2021acoustic,coutant2021topological,coutant2023topologically}.
It is worth noting that tight-binding models, usually arising in a regime of weak coupling in condensed matter, have found applications in systems with long range couplings such as elastic systems which exhibit a very strong coupling between different resonators as well as between various kind of elastic waves (transverse and longitudinal)~\cite{ramirez2020emulating,fan2022hermitian}.

The building blocks of discrete models are : 1) a set of sites where the relevant field is located and has a (complex) amplitude $A_n$ (where $n\in \mathbb{N}$ denotes the various sites), and 2) hopping coefficients which characterise the couplings between the different sites of the lattice. Those two quantities, the field amplitudes and the hopping coefficients, are quite generic but they can be related to physical quantities depending on the system of interest. For example, in the context of acoustic waves, the discrete model can be seen as an acoustic network of tubes where the field amplitude can be understood as the pressure field at the nodes of the network and the hopping coefficients are related to the different cross section areas of the tubes in the network~\cite{coutant2021acoustic}. In spring-mass networks, the amplitude is related to the displacements of the masses while the hoppings are linked to the spring constants and the masses~\cite{markos2008wave}, while in photonic crystals they can be related to the electric field amplitude and the dielectric constants of the system~\cite{vanel2017asymptotic}.

The consideration of lattice models allows us to solve the canonical scattering problem of valley Hall modes impinging on a sharp bend semi-analytically. Furthermore, the versatility of this approach will also allow for a comparison between our results and valley Hall scattering observations in a wide range of physical systems.



\subsection{2D periodic networks and the valley Hall effect}

\subsubsection{Graphene lattice and its band structure}
\begin{figure}
    \centering
    \includegraphics[scale=0.7]{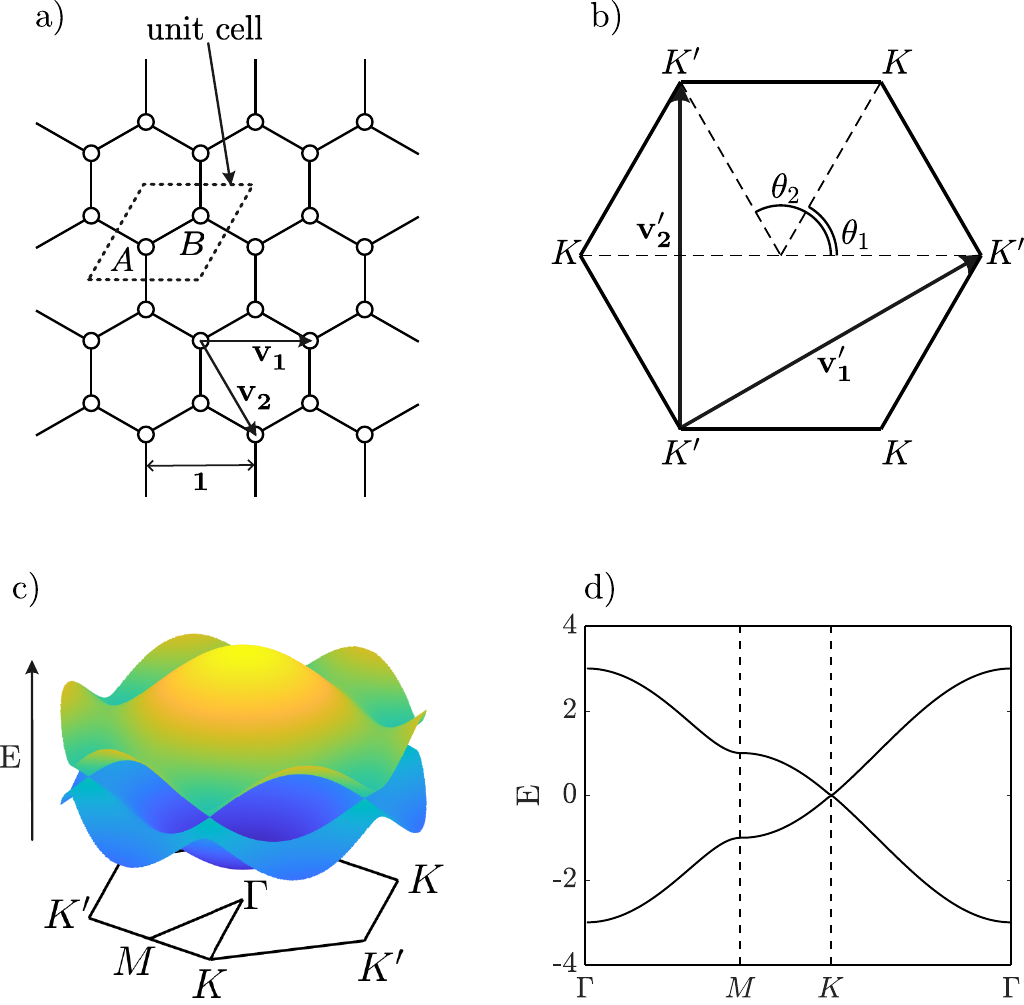}
    \caption{Graphene structure and its band structure. Panel a) depicts the graphene bulk structure, its unit cell and the lattice vectors $\vec{v_1} = (1,0)$ and $\vec{v_2} = (1/2,-\sqrt{3}/2)$. Panel b) represents its Brillouin zone with the $K$ and $K'$ Dirac points. $K/K'$ points are located at $\frac{4\pi}{3\sqrt{3}}(\pm 1,0)$ plus any integer shift by the reciprocal vectors \mbox{$\vec{v_1'} =\frac{4\pi}{\sqrt{3}}(\sqrt{3}/2,1/2)$} and \mbox{$\vec{v_2'} = 4\pi(1,0)$}. Panel c) depicts its band structure which meet at the Dirac points $K$ and $K'$. We also indicate the other high symmetry points $M$ and $\Gamma$. Panel d) shows the band structure along path that passes through the high symmetry points.}
    \label{fig:graphene_nogap_summary}
\end{figure}

We briefly review here the model and realisation of the valley Hall effect in graphene-like systems.
We begin by discussing the band properties of a discrete honeycomb lattice depicted in Fig.~\ref{fig:graphene_nogap_summary}-a). The lattice spacing is set to unity in the following without loss of generality.
The lattice is invariant under $\pi/3$ rotation, the 6 reflections of the hexagon  (i.e. its lattice point group is $C_{6v}$) and translations along the lattice vectors : $\vec{v_1} = (1,0)$ and $\vec{v_2}=(1/2,-\sqrt{3}/2)$. The entire lattice can be constructed from a unit cell with two sites, denoted A at position $\vec{R}_A$ and B at position $\vec{R}_B$, and shown in Fig.~\ref{fig:graphene_nogap_summary}-a).
Sites in the lattice can be located by two integers $p$ and $q$, and $A_{p,q}$ (respectively $B_{p,q}$) corresponds to the amplitude on the site at position $\vec{R}_A + p \vec{v_1} + q\vec{v_2}$ (respectively $\vec{R}_B + p \vec{v_1} + q\vec{v_2}$). 

Using the translation invariance, the amplitude on the sites of the lattice can be written in Bloch form as
\begin{eqnarray}
    A_{p,q} &=& e^{i \vec{k}.(p \vec{v_1} + q\vec{v_2})} A, \\
    B_{p,q} &=& e^{i \vec{k}.(p \vec{v_1} + q\vec{v_2})} B.
\end{eqnarray}
Introducing the Bloch Hamiltonian with Bloch wavenumber $\vec{k}$ as
\begin{equation}
    \vec{H}(\vec{k}) = \begin{pmatrix}
        0 & f(\vec{k}) \\
        f(\vec{k})^* & 0 \end{pmatrix}, 
        \quad 
        \text{where}
        \quad 
        f(\vec{k}) = 1 + e^{i\vec{k}.\vec{v_1}} + e^{-i\vec{k}.\vec{v_2}}
\end{equation}
and $f(\vec{k})^*$ is the complex conjugate to $f(\vec{k})$, the equations of motion reduce to
\begin{equation}
    \vec{H}(\vec{k}) \binom{A}{B} = E \binom{A}{B},
\end{equation}
where $E$ is an "energy" eigenvalue which relates to the frequency of the modes~\cite{vanel2017asymptotic,coutant2021acoustic}.
This Bloch Hamiltonian is a $2\times 2$ matrix with a hexagonal Brillouin zone depicted in Fig.~\ref{fig:graphene_nogap_summary}-b).
The two bands are shown in Fig.~\ref{fig:graphene_nogap_summary}-c) together with the high symmetry points $\Gamma$, $M$ and $K$ where the Bloch Hamiltonian acquires additional symmetry properties.
It can also be seen that the two bands meet in a conical shape at the Dirac points $K$ and $K'$. The neighbourhood of the Dirac points are called valleys. The band structure along a path passing through the high symmetry points is shown in Fig.~\ref{fig:graphene_nogap_summary}-d) where we clearly see that the gap closes at the $K$ point.

\subsubsection{Opening up the gap}
\begin{figure}
    \centering
    \includegraphics[scale=0.8]{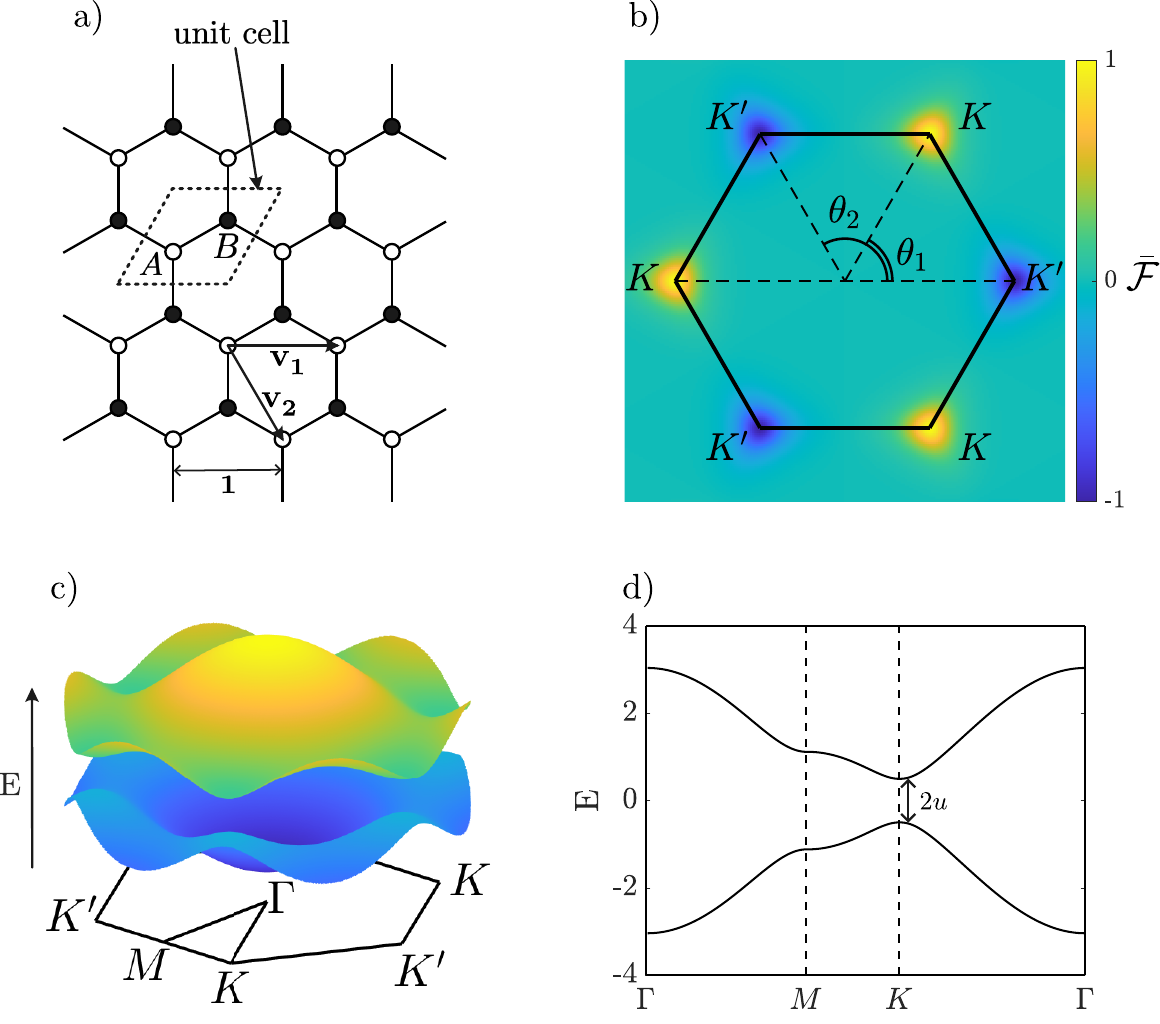}
    \caption{Graphene structure with symmetry breaking and illustration of the valley Hall effect with $u=0.5$. Panel (a) shows the graphene structure where we have broken a spatial symmetry. The black/white circles depict the sites with positive/negative onsite potential. Panel (b) depicts the normalised Berry curvature $\bar{\mathcal{F}}$ over the Brillouin zone which takes opposite signs in each valley. Panel (c) shows the two bands where the gap is clearly visible. Panel (d) shows the two energy bands along a path passing through the high symmetry points.}
    \label{fig:graphene_gap_summary}
\end{figure}

In order to realise the valley Hall effect, we now break a spatial symmetry of the system by introducing an onsite potential $u$, which takes opposite values on the A and B sites. 
This addition breaks inversion symmetry of the lattice which is now left invariant under $2\pi/3$ rotation, reflections along axes passing through vertices of the lattice and the translations along the vectors $\vec{a}$ and $\vec{b}$.
The lattice with broken spatial symmetry is shown in Fig.~\ref{fig:graphene_gap_summary}-a). We choose the black sites to have an onsite potential $+u$ and the white sites to have an onsite potential $-u$. In general $u$ will be chosen positive unless specified.

This modification of the lattice induces extra diagonal terms in the Bloch Hamiltonian which now takes the form 
\begin{equation}\label{HB_onsite}
    \vec{H}(\vec{k}) = \begin{pmatrix}
        u & f(k) \\
        f(k)^* & -u \end{pmatrix}.
\end{equation}
The energy eigenvalues are given by $E_\pm = \pm\sqrt{u^2 + |f|^2}$, which leads to a band structure presenting a gap of magnitude $2u$, as shown in Fig.~\ref{fig:graphene_gap_summary}-c) \& d).
While the energy eigenvalues are identical in the two valleys $K$ and $K'$, the modes in each valley are distinct. 
This becomes apparent when one computes the Berry curvature, shown in Fig.~\ref{fig:graphene_gap_summary}-b) (see supplemental material Sec.~\ref{App:Berry} for definition and computational details).
As one can see, the Berry curvature has opposite signs in the two valleys and it can be used as a pseudo-spin.
This construction together with arguments on the conservation of the valley index is at the basis of valleytronics and of the topological protection of valley Hall modes, which we shall investigate quantitatively in graphene ribbons in the next sections.

\subsection{Ribbon configurations}

After having discussed the band properties of infinite honeycomb lattices with and without onsite potential, we now turn our attention to the description of graphene ribbons with a finite width. 
As in the infinite lattice case, we will briefly review the properties of ribbons with and without onsite potential before constructing ribbons which support propagating valley states.

\subsubsection{Graphene ribbons}

When considering graphene ribbons, one has to decide how to ``cut'' the ribbons within the periodic graphene lattice. In other words, one has to choose the shape of the ribbon edges. There are infinitely many ways to cut the graphene lattice and authors have studied ribbons with various edges. 
Amongst all the possibilities stem three canonical choices: zigzag edges, bearded edges and armchair edges, which arise as straight cuts along high-symmetry directions of the honeycomb lattice~\cite{lado2022theory}. Zigzag and bearded edges are obtained when cutting along the direction of a lattice vector while the armchair is obtained when cutting in a direction perpendicular to a lattice vector.
Since the main focus of our work concerns modes localised near the interface and away from the edges, the direction of the ribbon interface is much more relevant that the edge type hence we will focus in the following on zigzag edges.
The case of armchair ribbons is discussed in the supplemental material Sec.~\ref{Sec:armchair}.

\begin{figure}
    \centering
    \includegraphics[scale=0.75]{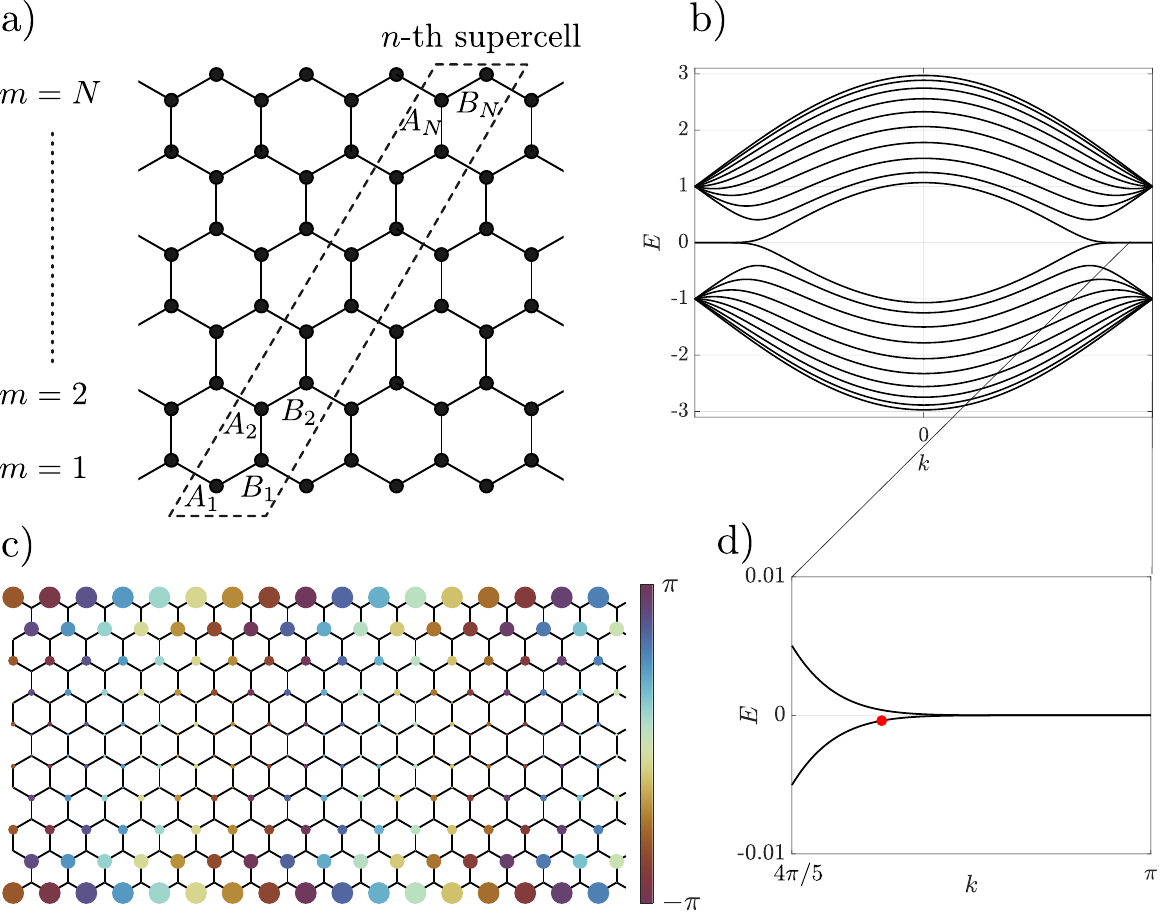}
    \caption{Band structure and edge modes of graphene ribbons with zigzag edges. Panel (a) shows a schematic of the graphene ribbon with zigzag edge. Panel (b) is the dispersion relation for a ribbon with N=10. 
    Panel (c) represents an eigenmode for a particular value of $k$ and $E$ indicated by the red dot in Panel (d). The area of the dots represent the amplitude of the mode while the color illustrates the phase.
    In addition, Panel (d) shows a close up version on the nearly flat bands near $k = \pi$. }
    \label{fig:graphene_ribbon_summary}
\end{figure}

We consider a graphene ribbon with zigzag edges, as depicted in Fig.~\ref{fig:graphene_ribbon_summary}-a).
The ribbon is invariant under translation along the vector $\vec{v}_1$, and can therefore be constructed by translating the supercell containing $2N$ sites delimited by the dashed box in Fig.~\ref{fig:graphene_ribbon_summary}-a). The particular choice of a tilted supercell is made as it will simplify the equations of motion compared to the more classical choice of a straight supercell~\cite{wakabayashi2010electronic}.
The supercell can itself be constructed from a unit cell identical to the one of the graphene lattice of Fig.~\ref{fig:graphene_nogap_summary}-a). A and B sites in the ribbons can be identified by two integers $m\in [1,N]$ and $n\in \mathbb{Z}$. $A_{m,n}$ and $B_{m,n}$ are the complex amplitudes on the A and B sites in $m$-th row of the $n$-th supercell.
The equations describing the amplitudes on the various sites of the ribbon are 
\begin{eqnarray}
    E A_{m,n} &=& B_{m-1,n} + B_{m,n} + B_{m,n-1}, \label{eq:mn_eq1}\\
    E B_{m,n} &=& A_{m+1,n} + A_{m,n} + A_{m,n+1},\label{eq:mn_eq2}
\end{eqnarray}
defining the ghost sites $B_{0,n}=A_{N+1,n} = 0$, these equations are valid for $1\leq m\leq N$ and \mbox{$n\in \mathbb{Z}$}.
Using the translational invariance, we can write the amplitudes in Bloch form with Bloch wavenumber $k$
\begin{equation}
    A_{m,n} = e^{ikn}A_m \quad \text{and} \quad B_{m,n} = e^{ikn}B_m.
\end{equation}
The equations of motion are given by
\begin{eqnarray}
    EA_{m} &=& B_{m-1} + B_{m} + e^{-ik}B_{m}, \\
    EB_{m} &=& A_{m+1} + A_{m} + e^{ik}A_{m},
\end{eqnarray}
together with the definition of the ghost sites $B_0=A_{N+1} = 0$
which are valid for $1\leq m\leq N$.
Introducing the parameter
\begin{equation}
    s(k) = 2e^{-ik/2}\cos(k/2),
\end{equation}
the governing equations take the form
\begin{eqnarray}\label{SSH1}
    EA_{m} &=& B_{m-1} + s(k)B_{m}, \\
    EB_{m} &=& A_{m+1} + s(k)^*A_{m}.
\end{eqnarray}
These equations can be written as an eigenvalue problem for the Bloch Hamiltonian $H(k)$ as
\begin{equation}
    \vec{H}(k)\Psi = E\Psi, \quad \text{with} \quad \Psi = \begin{pmatrix}
        A_1,
        B_1,
        \cdots,
        A_N,
        B_N
    \end{pmatrix}^T.
\end{equation}
Explicitly, the Bloch Hamiltonian is given by
\begin{equation}
    \vec{H}(k) = \begin{pmatrix}
        0 & s & 0 & \cdots & \cdots & \cdots &0  \\
        s^* & 0 & 1 & 0 & & & \vdots \\
        0 & 1 & 0 & s & \ddots & & \vdots \\
        \vdots & 0 & s^* & 0 & 1 & \ddots & \vdots \\
        \vdots & & \ddots & \ddots & \ddots &\ddots & 0 \\
        \vdots & &  & \ddots & \ddots & \ddots & s \\
        0 & \cdots  &\cdots  &\cdots  &0 &s^* & 0 
    \end{pmatrix}
\end{equation}
The Bloch Hamiltonian is hermitian, hence its spectrum is real.
We recognize here the well-known SSH model~\cite{su1979solitons,asboth2016short} with a $k$-dependent complex intracell hopping parameter (the intercell hopping is equal to unity). 
Note that the standard SSH model is usually defined with real hopping coefficients. It is possible to reduce our system to a SSH with real hoppings by defining a local phase transformation of the amplitudes 
according to
\begin{equation}\label{eq:change_of_var1}
    \tilde{A}_m = \begin{cases}
A_m & \text{if } m \text{ is even}\\
e^{-ik/2}A_m & \text{if } m \text{ is odd},
\end{cases}
\end{equation}
and
\begin{equation}\label{eq:change_of_var2}
\tilde{B}_m = \begin{cases}
e^{-ik/2}B_m & \text{if } m \text{ is even}\\
B_m & \text{if } m \text{ is odd}.
\end{cases}
\end{equation}
One can therefore construct the eigenmodes of the graphene ribbon with those of the SSH chain via this phase redefinition while the spectrum of the graphene ribbon is not affected by this unitary transformation.
Therefore at fixed Bloch wavenumber $k$, the energy eigenvalues of the graphene ribbon supercell are those of a SSH chain with $2N$ sites and intracell hoping $s(k)$.
The dispersion relation of the graphene ribbon with $N=10$ is shown in Fig.~\ref{fig:graphene_ribbon_summary}-b). 
The band structure can be separated into 3 parts : 2 bulk parts which contain the majority of the bands and a set of two isolated bands which become nearly flat near $E=0$ for $k\sim\pm\pi$ (see Fig.~\ref{fig:graphene_ribbon_summary}-d)). 
For $|k|\gtrsim 2\pi/3 $, those two bands correspond to edge states of the ribbon~\cite{wakabayashi2010electronic}. 
This can be understood in terms of the SSH chain, since the intracell hopping $|s(k)|<1$ for $k>2\pi/3$ and it therefore corresponds to a topological phase of the SSH chain. 
One such edge mode corresponding to the energy indicated by the red dot in Fig.~\ref{fig:graphene_ribbon_summary}-d) is depicted in Fig.~\ref{fig:graphene_ribbon_summary}-c).
The bulk part contains modes which correspond to the transverse modes of the ribbon and are not localised near the edges.

It is possible to characterise further the edge modes of the graphene ribbon by analysing the symmetry properties of the ribbon and the graphene structure.

From the projector operators, $\vec{P_A}$ and $\vec{P_B}$, defined as $\vec{P_A}\Psi=(A_1,0,A_2,0,\cdots,A_N,0)^T$ and $\vec{P_B}\Psi=(0,B_1,0,B_2,\cdots,0,B_N)^T$, which projects the amplitudes on the A and B sublattices respectively, we define the chiral operator 
\begin{equation}\label{eq:Gamma}
    \vec{\Gamma} = \vec{P}_A - \vec{P}_B.
\end{equation}
In other words, the chiral operator leaves the amplitude on the A sites identical but flips the sign of the amplitude on the B sites.
Since the lattice does not contain any coupling between two sites of the A and B sublattice, we have that its Hamiltonian satisfies
\begin{equation}
    \bm{\Gamma} \vec{H} \bm{\Gamma} = - \vec{H}.
\end{equation}
This sublattice symmetry, also called chiral symmetry~\cite{asboth2016short}, implies in particular that if $v$ is an eigenvector associated to the energy $E$, then $\vec{\Gamma} v$ is an eigenvector associated with energy $-E$. Therefore the energy spectrum is symmetric across $E=0$. 
Furthermore, using the mirror symmetry of the graphene ribbon along the horizontal axis, one can decompose its eigenmodes into symmetric and antisymmetric sectors. 
In particular, edge modes on both sides split into symmetric and antisymmetric edge modes which are localised on both edges of the ribbons simultaneously, as shown in Fig.~\ref{fig:graphene_ribbon_summary}-c).

\subsubsection{Graphene ribbons with onsite potential}

\begin{figure}
    \centering
    \includegraphics[scale=1]{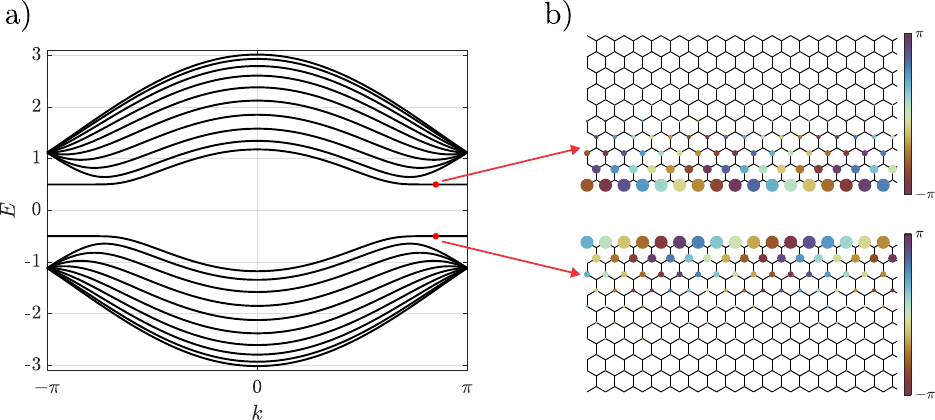}
    \caption{Band structure and edge modes of graphene ribbons with zigzag edges and onsite potential $u=0.5$. Panel~(a) shows the dispersion relation which exhibits a band gap. Panel~(b) shows two edge modes located either on the lower or the upper edge of the ribbon. The area of the dots represents the amplitude of the mode while the color illustrates the phase.}
    \label{fig:graphene_ribbon_onsite_summary}
\end{figure}

The graphene ribbon described above does not exhibit any band gap.
As for the infinite lattice, we now introduce an onsite potential to open a gap around the energy $E=0$.
\\
We consider a graphene ribbon with zigzag edges, with the addition of an onsite potential $+u$ on the A sites and $-u$ on the B sites of the lattice.
The governing equations become 
\begin{eqnarray}\label{RM1}
    EA_{m} &=& u A_m + B_{m-1} + sB_{m}, \\
    EB_{m} &=& -u B_m +A_{m+1} + s^*A_{m}.
\end{eqnarray}
These describe an extension to the SSH chain, known as the Rice-Mele model~\cite{asboth2016short}.
Therefore the band structure of the graphene ribbon with an onsite potential can be obtained from the spectrum of a Rice-Mele chain.
In particular, it will exhibit a band gap of size $2u$ and allow for the presence of edge modes, which are shown in Fig.~\ref{fig:graphene_ribbon_onsite_summary}.
Fig.~\ref{fig:graphene_ribbon_onsite_summary}-a) shows the dispersion relation of the graphene ribbon with zigzag edges, $N=10$ and an onsite potential $u=0.5$. The band gap is visible around the energy $E=0$. Fig.~\ref{fig:graphene_ribbon_onsite_summary}-b) illustrates an edge modes for the energy and wavenumber corresponding to the red dot in Panel (a).
Note that the edge modes are now localised on one edge of the ribbon, since the addition of the onsite potential broke the mirror symmetry of the ribbon along the horizontal axis. 

As can be seen from Fig.~\ref{fig:graphene_ribbon_onsite_summary}-a), the addition of the onsite potential opened a gap around $E=0$. We have shown previously that in the bulk of the ribbon, the addition of an onsite potential leads to distinct valley modes at $K$ and $K'$ equipped with a valley index. To see these modes manifest in the ribbon and create a topological insulator, we now need to connect two ribbons in opposite topological phases. 
That is we need to create an interface between two ribbons with opposite onsite potential, which we will study in the next section.

\subsubsection{Graphene ribbons with an interface}\label{subsec:interface}

\begin{figure}
    \centering
    \includegraphics[scale=0.8]{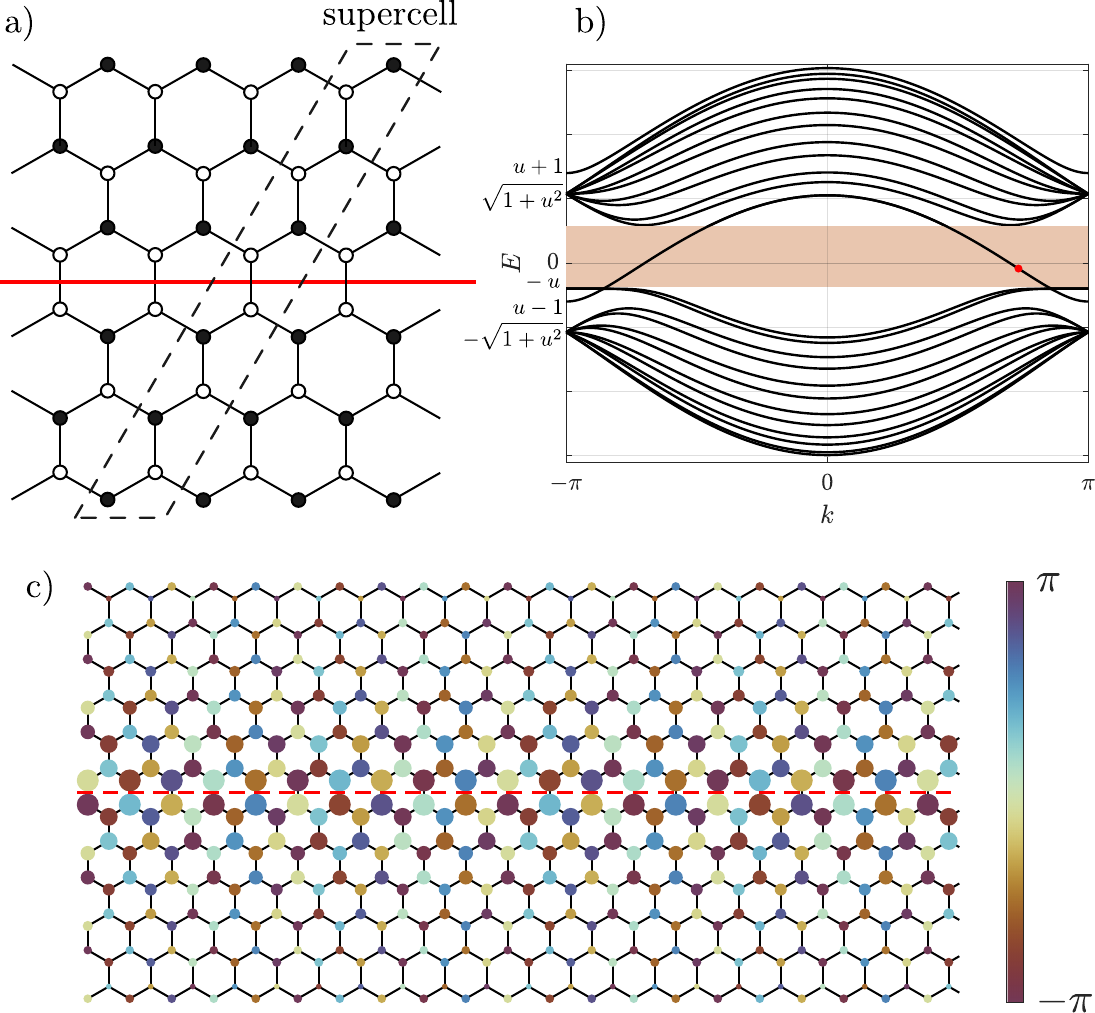}
    \caption{Band structure and interface modes of graphene ribbons with a bridge interface. Panel (a) shows the geometrical structure of the ribbon and the interface. Panel (b) depicts the band structure of a ribbon with bridge interface with $N=12$ and $u=0.4$. The red shaded region indicates the energy range with a single pair of propagating modes.
    Panel (c) illustrates an interface mode corresponding to the energy shown with the red dot in Panel (b).}
    \label{fig:ribbons_bridge}
\end{figure}

\begin{figure}
    \centering
    \includegraphics[scale=0.8]{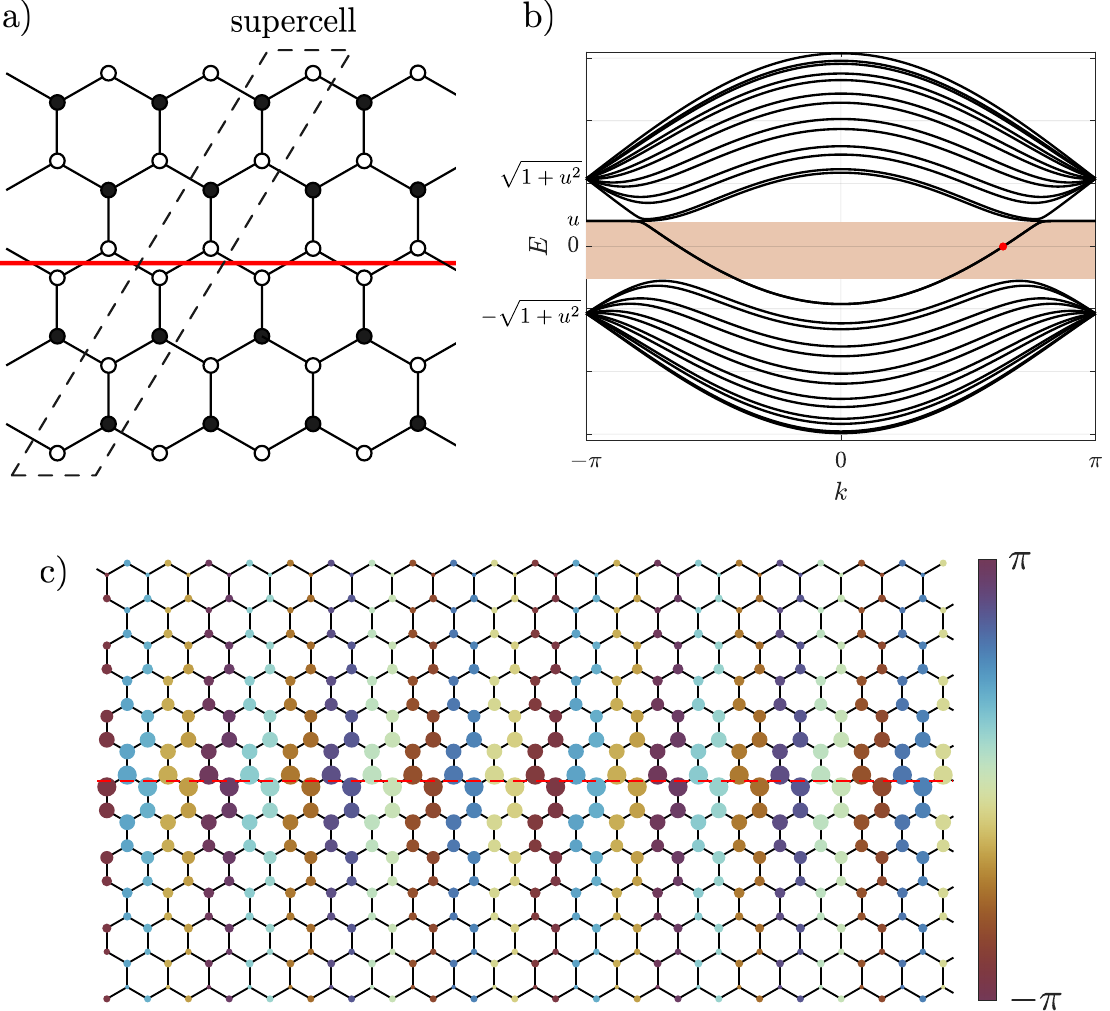}
    \caption{Band structure and interface modes of graphene ribbons with a zigzag interface. Panel (a) shows the geometrical structure of the ribbon and the interface. Panel (b) depicts the band structure of a ribbon with bridge interface with $N=13$ and $u=0.4$. 
    The red shaded region indicates the energy range with a single pair of propagating modes.
    Panel (c) illustrates an interface mode corresponding to the energy shown with the red dot in Panel (b).}
    \label{fig:ribbons_zigzag}
\end{figure}

We now consider graphene ribbons with the addition of an onsite potential and in the presence of an interface. 
The interface will always be considered in the middle of the ribbon hence there are two possible configurations for the interface : i) a bridge interface or ii) a zigzag interface.
The ribbons are made of supercells, shown in dotted rectangles in Figs.~\ref{fig:ribbons_bridge} and~\ref{fig:ribbons_zigzag}, containing each $N$ pairs of sites. 
We choose to put the interface in the middle of the ribbon width, hence if $N$ is even the interface will be a bridge interface (Fig.~\ref{fig:ribbons_bridge}) while for odd values $N$ the interface will be of zigzag type (Fig.~\ref{fig:ribbons_zigzag}).
Furthermore, without loss of generality, the value of $u$ is chosen such that the nearest sites to the interface have a positive onsite potential (the case of a negative $u$ at the interface is discussed at the end of the section).
As was done previously, each site in the ribbon can be localised using two integers $(m,n)$, where $n$ corresponds to the supercell index given an arbitrary origin and $m$ indicates the position of the site in the supercell.

Again, the problem is invariant under translation along the direction of the ribbon, hence we can look for eigenmodes with Bloch wavenumber $k$. 
The governing equations are of the form of Eqs.~\eqref{RM1} with the appropriate sign for the onsite potential depending on the side of the interface. These equations can also be more conveniently written, using the Bloch Hamiltonian of the supercell $H(k)$ and the vector $\psi$ containing the amplitudes on the $2N$ sites of the supercell, as an eigenvalue problem of the form
\begin{equation}\label{eq:Bloch_ribbon}
    \vec{H}(k) \psi = E \psi.
\end{equation}
At fixed $k$ there are $2N$ eigenmodes propagating along the ribbon with momentum $k$.
Since $\vec{H}(k)$ is Hermitian, its spectrum is real. 
The dispersion relations are obtained by varying $k$ in the 1-dimensional Brillouin zone and diagonalizing $\vec{H}(k)$ for each different Bloch momenta.

The dispersion relation for graphene ribbon with the two possible interfaces are represented in Figs.~\ref{fig:ribbons_bridge}-b) and~\ref{fig:ribbons_zigzag}-b). It can be clearly seen that the dispersion relation presents a band gap and that there is a single band which bridges the two bulk parts. 
This band corresponds to an interface state localised near the interface. 
Furthermore, the modes present in the ribbon can be separated in three sets : a pair of edge modes, a pair of interface modes and bulk modes. 

This separation becomes apparent at the edges of the Brillouin zone, where we can distinguish a set of special energy values depending on the type of interface considered. 
Those energy values can be extracted from the limit $s(k) = 0$, corresponding to the fully dimmerized Rice-Mele chain, and calculations are presented in the supplemental material Sec.~\ref{App:RM}.
When $k \rightarrow \pm \pi$ and therefore $s(k)\rightarrow 0$, we can see that the bulk modes accumulate around the energies $E=\pm \sqrt{1 + u^2}$. For a bridge interface the edges modes have energy $E=-u$ and the interface modes have energies $E= u \pm 1$. For a zigzag interface the edges modes have energy $E=u$ and the interface modes have energies $E=\sqrt{1+u^2}$ and join the bulk modes.

We further note that close to the edge of the Brillouin zone, the interface band meets the edge state band at $\pm u$, this implies that the associated eigenvectors will be less localised near the interface but will spread out across ribbon until they relocalize near the edge of the ribbon as $k \rightarrow \pm \pi$. Similarly, around $k = 0$, the interface band joins the bulk and interface states again spread away from the interface.

Figs.~\ref{fig:ribbons_bridge}-c) and~\ref{fig:ribbons_zigzag}-c) show interface states in both configurations. 
Note that the addition of an interface restores the mirror symmetry of the Rice-Mele chain associated with the ribbon supercell.
The eigenmodes of the chain (and therefore of the ribbon provided the phase transformation given in Eqs.~\eqref{eq:change_of_var1} \&~\eqref{eq:change_of_var2}) can again be decomposed into a symmetric and antisymmetric sector.
It can be shown that the interface mode in the band gap is an antisymmetric mode.
The energy and symmetry properties of the interface states can be understood in the limit of ribbons with infinite width which is presented in Sec.~\ref{App:RM} of the supplemental material.

We have presented here the spectrum of the various interface configurations for a fixed value of $u=0.4$. 
Ribbons with the opposite sign of $u$ are also possible, and will indeed be relevant in the following scattering scenarios considered below. 
We note that ribbons with onsite potential $u$ and $-u$ are related via the simple transformation $E \rightarrow -E$. Indeed, suppose that $v$ is an eigenvector Eq.~\eqref{eq:Bloch_ribbon} with the onsite potential $u$.
Denoting the components of $v$ on the A/B sites as $v^{A/B}$, Eq.~\eqref{eq:Bloch_ribbon} can be explicitly written in the form Eq.~\eqref{RM1} as 
\begin{eqnarray}
    (E - \vec{U}_A) v^A_m &=& v^B_{m-1} + sv^B_m \\
    (E - \vec{U}_B) v^B_m &=& v^A_{m+1} + s^* v^A_m,
\end{eqnarray}
where $U_{A/B}$ are diagonal matrices containing the values of the onsite potential on the A/B sites.
These equation can be rewritten as 
\begin{eqnarray}
    (-E + \vec{U}_A) \ v^A_m &=& -v^B_{m-1} - sv^B_m \\
    -(-E + \vec{U}_B)\ v^B_m &=& v^A_{m+1} + s^* v^A_m.
\end{eqnarray}
In other words, the vector $\bm{\Gamma} v = (v^A,-v^B)$ is an eigenvector of Eq.~\eqref{eq:Bloch_ribbon} with energy $-E$ and onsite potential $-u$, where $\vec{\Gamma}$ is the chiral operator introduced previously.


In the following we will be interested in the scattering of valley modes, hence we will focus on energy ranges where only the interface modes are propagating. This range depends on both $u$ and $N$ and is not symmetric around $E=0$. Those energy ranges, for the specific values of $u=0.4$ and $N=12$ and $13$, are shown as the light red shaded regions in Figs.~\ref{fig:ribbons_bridge}-b) and~\ref{fig:ribbons_zigzag}-b).

\section{Transfer matrix formalism}\label{sec:transfer}

In the above section, we have derived the band structure of graphene ribbon with different interface types using the translational invariance of the ribbon and Bloch's theorem. 
In particular we have characterised the propagating modes in the ribbon and exhibited the presence of interface states for graphene ribbons with zigzag edges and both bridge and zigzag interface.
We have shown that there exists an energy range for which only the interface state can propagate in the ribbon.
Our goal is to study the scattering of these interface modes across bends in this energy range.
To achieve this, we require knowledge of both the propagating interface states and the evanescent modes that can be excited through the scattering.
Here, we develop a transfer matrix formalism which will allow us to characterise both the propagating and evanescent modes simultaneously and eventually quantify the scattering properties in the canonical scenarios of interest.
We begin by constructing the transfer matrix formalism for the graphene ribbon before extending it to incorporate bends and other scenarios.

\subsection{Transfer matrix of ribbons}\label{sec:TM_ribbons}

We begin our discussion of the transfer matrix formalism~\cite{dwivedi2016bulk,coutant2023topologically} by focusing on the case of graphene ribbons with zigzag edges. 
This will allow us to introduce key ingredients in a simplified case before turning our attention to scattering problems. 
It will also provide an alternative way to construct the dispersion relation and to access the evanescent modes in the ribbon.

\begin{figure}
    \centering
    \includegraphics[scale = 1]{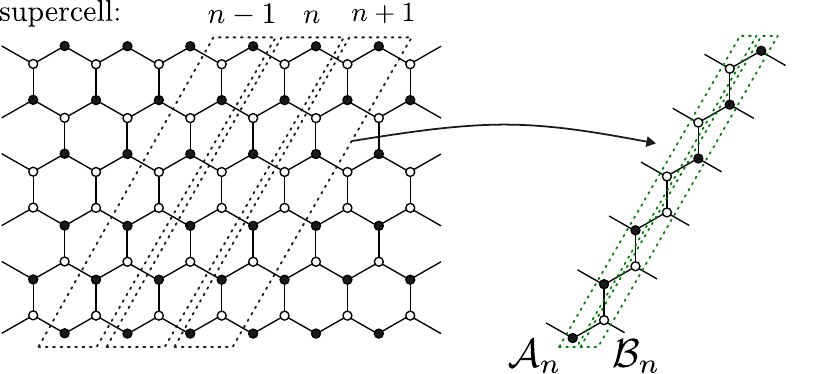}
    \caption{Illustration of the supercells of the graphene ribbons with zigzag edges and a bridge interface together with their subdivision into the site sets and amplitudes $\mathcal{A}_n$ $\mathcal{B}_n$.}
    \label{fig:transfer_ribbon_cell}
\end{figure}

As was done previously we define the supercell of the ribbon and introduce the vector $\mathcal{A}_n$ and $\mathcal{B}_n$ containing the amplitudes on the A and B sites of the supercell $n$ (see Fig.~\ref{fig:transfer_ribbon_cell}).

Explicitly, these vectors are
\begin{equation}
    \mathcal{A}_n = (A_{1,n},A_{2,n},\cdots,A_{N,n})^T \quad \text{and} \quad \mathcal{B}_n = (B_{1,n},B_{2,n},\cdots ,B_{N,n})^T.
\end{equation}
We introduce the matrix $\vec{J}$ defined as 
\begin{equation}
 \vec{J} = 
 \begin{pmatrix}
     0 & \cdots & \cdots & \cdots & 0 \\
     1 & \ddots & & & \vdots \\
     0 & \ddots & \ddots & & \vdots \\
     \vdots & \ddots & \ddots & \ddots & \vdots \\
      0 & \cdots & 0 & 1 & 0
 \end{pmatrix}
 \end{equation}
as well as the diagonal matrices $\vec{U}_A$ and $\vec{U}_B$ whose diagonal elements $(\vec{U}_A)_{ii}$ and $(\vec{U}_B)_{ii}$ are the value of the onsite potential on the sites $A_{i,n}$ and $B_{i,n}$ (these values do not depend on the supercell index $n$). 
We note that $\vec{U}_A = -\vec{U}_B$. 
With these matrices, the governing equations Eqs.~\eqref{eq:mn_eq1} and~\eqref{eq:mn_eq2}, with the addition of the onsite potential can be compactly written as
\begin{eqnarray}
    E \mathcal{A}_n &=& \vec{U}_A \mathcal{A}_n + \vec{J} \mathcal{B}_n + \mathcal{B}_n + \mathcal{B}_{n-1},  \\
    E \mathcal{B}_n &=& \vec{U}_B \mathcal{B}_n + \vec{J}^T \mathcal{A}_n + \mathcal{A}_n + \mathcal{A}_{n+1}.
\end{eqnarray}
These can be rearranged as
\begin{eqnarray}
    \mathcal{A}_{n+1} + \left( \vec{U}_B - E\ \id_N \right) \mathcal{B}_n &=& -\left( \id_N + \vec{J}^T\right) \mathcal{A}_n \label{eq:AnBn1}\\
    \left( \id_N + \vec{J} \right) \mathcal{B}_n &=& \left( E\ \id_N - \vec{U}_A\right) \mathcal{A}_n - \mathcal{B}_{n-1},\label{eq:AnBn2}
\end{eqnarray}
where $\id_N$ is the $N\times N$ identity matrix.
We now introduce the vector $\Psi_m$ and the two matrices $\vec{M}_1$ and $\vec{M}_2$ defined as
\begin{equation}
\Psi_n = \binom{\mathcal{A}_n} {\mathcal{B}_{n-1}}, \quad
\vec{M}_1 = \begin{pmatrix}
    \id_N & \vec{U}_B - E\ \id_N \\
    \bm{0}_N & \id_N + \vec{J}
\end{pmatrix}, \quad \text{and} \quad
\vec{M}_2 = \begin{pmatrix}
    -\left(\id_N + \vec{J}^T \right) & \bm{0}_N \\
     (E\ \id_N - \vec{U}_A)& - \id_N 
\end{pmatrix}.
\end{equation}
With these quantities, the governing equation are given by
\begin{eqnarray}\label{eq:gen_rec}
    \vec{M}_1 \Psi_{n+1} = \vec{M}_2 \Psi_n, \quad \text{with}\quad \Psi_n \in \mathbb{C}^{2N}
\end{eqnarray}
From the block decomposition of $\vec{M}_1$, we can see that $\vec{M}_1$ is invertible since 
it is block triangular and $\id_N + \vec{J}$ is invertible. 
We can therefore invert the matrix $\vec{M}_1$ and construct the transfer matrix of the ribbon, $\vec{M}$ , to simplify Eq.~\eqref{eq:gen_rec} as
\begin{equation}
    \Psi_{n+1} = \vec{M} \Psi_n,\ \text{with} \ \ \vec{M}=\vec{M}_1^{-1} \vec{M}_2.
\end{equation}

\subsection{Properties of the transfer matrix}\label{subsec:properties}
\subsubsection{Conserved current}

Due to the fact that the Hamiltonian of an isolated supercell of the ribbon is self-adjoint, there exists a conserved current in the ribbon. 
This can most easily be seen from Eqs.~\eqref{eq:AnBn1} and~\eqref{eq:AnBn2}. By multiplying Eq.~\eqref{eq:AnBn1} by $\mathcal{B}_n^\dagger$ and Eq.~\eqref{eq:AnBn2} by $\mathcal{A}_n^\dagger$, we get
\begin{eqnarray}
    \mathcal{B}_n^\dagger \mathcal{A}_{n+1} + \mathcal{B}_n^\dagger(\vec{U}_B - E\id_N)\mathcal{B}_n &=& - \mathcal{B}_n^\dagger\left(\id_N + \vec{J}^T \right) \mathcal{A}_n \\
    \mathcal{A}_n^\dagger \left( \id_N + \vec{J} \right) \mathcal{B}_n &=& \mathcal{A}_n^\dagger \left( E \id_N - \vec{U}_A \right) \mathcal{A}_n - \mathcal{A}_n^\dagger\mathcal{B}_{n-1},
\end{eqnarray}
where $\dagger$ represents the Hermitian transpose operation.
Using the fact  that $(E\id_N - \vec{U}_{A/B})$ is a diagonal matrix, we can combine the two above expressions to get
\begin{equation}
   \mathcal{B}_n^\dagger \mathcal{A}_{n+1} - \mathcal{B}_{n-1}^\dagger \mathcal{A}_{n} = - \mathcal{B}_n^\dagger(\vec{U}_B - E\id_N) \mathcal{B}_n + \mathcal{A}_n^\dagger(\vec{U}_A - E\id_N) \mathcal{A}_n \in \mathbb{R}.
\end{equation}
Defining 
\begin{equation}
    \mathcal{J}_n = \mathrm{Im}\left(\mathcal{B}_{n-1}^\dagger \cdot\mathcal{A}_{n}\right),
\end{equation}
we have
\begin{equation}
    \mathcal{J}_{n+1} = \mathcal{J}_n,
\end{equation}
and hence $\mathcal{J}_n$ is conserved along the ribbon and can be used to normalise the propagating modes.

Note that $\mathcal{J}_n$ can be conveniently written as 
\begin{equation}
    \mathcal{J}_n = \frac{1}{2}\binom{\mathcal{A}_n}{\mathcal{B}_{n-1}}^\dagger\cdot \begin{pmatrix}
    \bm{0}_N & i\id_N \\
    -i \id_N & \bm{0}_N
\end{pmatrix} \cdot \binom{\mathcal{A}_n}{\mathcal{B}_{n-1}}.
\end{equation}
The conservation of $\mathcal{J}_n$ along the ribbon, implies that the transfer matrix $M$ is symplectic, i.e. it satisfies
\begin{equation}
    \vec{M}^\dagger \cdot \begin{pmatrix}
    \bm{0}_{N} & \id_{N} \\
     \id_{N} & \bm{0}_{N}
\end{pmatrix} \cdot \vec{M} = \begin{pmatrix}
    \bm{0}_{N} & \id_N \\
     \id_{N} & \bm{0}_N
\end{pmatrix}.
\end{equation}

\subsubsection{Eigenmodes}
The modes of the ribbon are given by the eigenvectors of the transfer matrix,
\begin{equation}
    \vec{M}\psi_j = \lambda_j \psi_j,
\end{equation}
with $j=1,..., N$. Writing the eigenvalues as $\lambda_j = e^{ik_j}$, we see that if $|\lambda_j| = 1$ the mode is propagative and if $|\lambda_j| \neq 1$, the mode is evanescent. For propagating modes, $k_j$ is real and it corresponds to the Bloch wavenumber. The wavevector associated to the eigenvalue $\lambda_j$ is denoted $\psi_j$.
Since the transfer matrix is symplectic, eigenvalues come in reciprocal pairs $(\lambda_j,\lambda_j^{-1})$. Half of the modes are right moving and the other half is left moving.
The direction is determined by the modulus of $\lambda_j$: if $|\lambda_j|<1$ the mode is moving to the right, while if $|\lambda_j|>1$ the mode is moving to the left. If $\lambda_j$ is of unit modulus, there are two options to determine its direction of propagation. One can compute the group velocity $v_g = (\partial_E k_j)^{-1}$, or one can add a small fictitious dissipation $E \rightarrow E + i \nu$ with $\nu > 0$ and apply the previous modulus criterion.
In the following we adopt the later approach and we will use $\nu = 10^{-10}$.

To prepare for the scattering scenarios, it is convenient to gather the eigenmodes and eigenvalues into matrix form
\begin{equation}
    \vec{M}\cdot\begin{pmatrix}\vec{A}\\ \vec{B}\end{pmatrix}= \vec{D}\cdot\begin{pmatrix}\vec{A}\\ \vec{B}\end{pmatrix},
\end{equation}
where $\vec{D} = \text{diag}(\lambda_j)$, $\vec{A}_{ij} = ((\bm{1}_N \ \bm{0}_N)\cdot  \psi_j)_i$ and $\vec{B}_{ij} = ((\bm{0}_N \ \id_N)\cdot \psi_j)_i$. We can further decompose $\vec{A}$ and $\vec{B}$ restricting their column to right and left moving propagating modes. Doing so, we define the matrices $\vec{A}^\pm$, $\vec{B}^\pm$ and $\vec{D}^\pm$ associated with right moving (+) and left moving (-) modes, $\psi^+$ and $\psi^-$ respectively.
Furthermore, the conserved current defined above can be used to normalize the propagating modes such that their current is $\pm 1$.

\subsection{Application to scattering}\label{sec:TM_bend}

\begin{figure}
    \centering
\includegraphics[scale = 0.8]{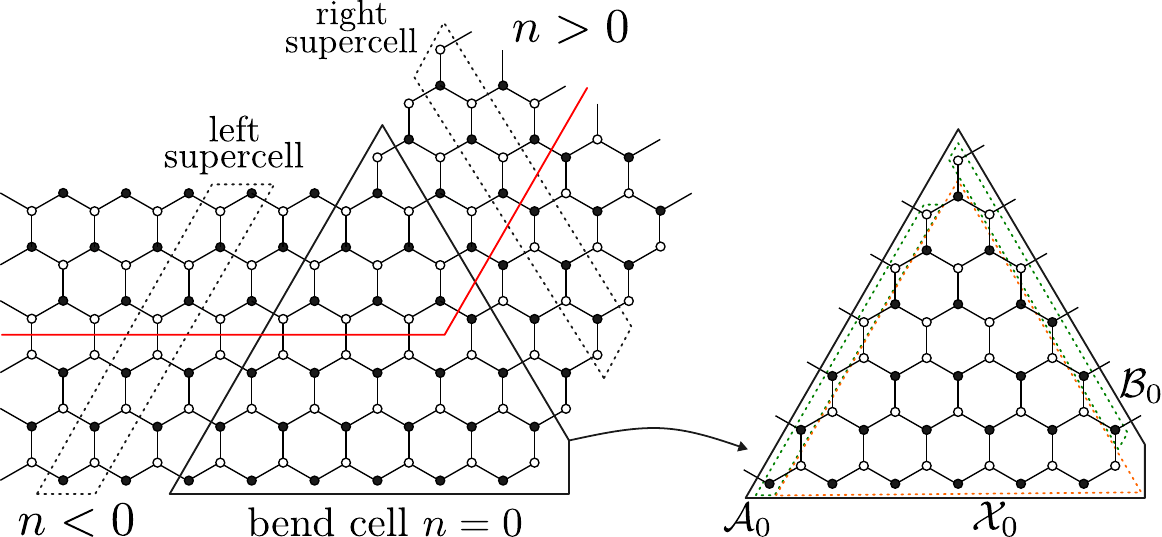}
    \caption{Illustration of the various supercells in a ribbon configuration with a bridge interface and a bend of angle $\pi/3$.}
    \label{fig:defect}
\end{figure}

In the previous section, we have constructed the forward and backward propagating modes which we now use to define scattering quantities through a bend connecting two semi-infinite periodic ribbons. This situation is represented schematically for a specific choice of bend and of interface in Fig.~\ref{fig:defect}.

We can label supercells in the left ribbon with the integer $n<0$ and in the right ribbon with $n>0$. 
The bend is located in an isolated cell at $n=0$. An important point in defining the bend cell is that the site on the left (right) edge of the defect cell must have the same arrangement than the ones on the left (right) supercell of the ribbons.

\subsubsection{Connection across a bend}
In the previous section, we have constructed the propagating and evanescent eigenmodes and their associated eigenvalues in a ribbon configuration.
We now need to connect those modes on each side of the bend. 
To do so, we will construct the transfer matrix of the bend cell. 

The first step is to introduce the amplitudes on three complementary sets of sites, $\mathcal{A}_0$, $\mathcal{B}_0$ and  $\mathcal{X}_0$ (see Fig.~\ref{fig:defect}), such that:
\begin{itemize}
    \item $\mathcal{A}_0$ corresponds to the amplitude on the sites which belong to the left edge of the bend cell $n=0$ and are connected to sites of the left supercell $n=-1$.
    \item $\mathcal{B}_0$ corresponds to the amplitude on the sites which belong to the right edge of the bend cell $n=0$ and are connected to sites of the right supercell $n=1$.
    \item $\mathcal{X}_0$ corresponds to the amplitude on the sites of the bend cell $n=0$ which are neither connected to the left nor the right ribbon.
\end{itemize}
We further denote $n_0$ the number of sites in the vectors $\mathcal{A}_0$ and $\mathcal{B}_0$, and $N$ the total number of site in the bend cell at $n=0$.
To these sets of sites, we define their associated projector $\vec{P}_\mathcal{A}$, $\vec{P}_\mathcal{B}$ and $\vec{P}_\mathcal{X}$ to account for the extra sites
\begin{equation}\label{eq:projectors}
    \vec{P}_\mathcal{A} = \begin{pmatrix}
        \bm{1}_{n_0 \times n_0} \\
        \bm{0}_{n_0 \times n_0} \\
        \bm{0}_{(N-2n_0) \times n_0}
    \end{pmatrix}, \quad
    \vec{P}_\mathcal{B} = \begin{pmatrix}
        \bm{0}_{n_0 \times n_0} \\
        \bm{1}_{n_0 \times n_0} \\
        \bm{0}_{(N-2n_0) \times n_0}
    \end{pmatrix},\quad\text{and} \quad
    \vec{P}_\mathcal{X} = \begin{pmatrix}
        \bm{0}_{2n_0 \times (N-2n_0)} \\
        \bm{1}_{(N-2n_0) \times n_0}
        \end{pmatrix}.
\end{equation}
For the projectors to have the simple form above, we have chosen to order and label the sites within the bend cell as $\Phi_0 = \left( \mathcal{A}_0,\mathcal{B}_0,\mathcal{X}_0 \right)^T$.
With the notation above, the amplitude in the bend can be decomposed as
\begin{equation}
    \Phi_0 = \vec{P}_\mathcal{A} \mathcal{A}_0 + \vec{P}_\mathcal{B} \mathcal{B}_0 + \vec{P}_\mathcal{X} \mathcal{X}_0.
\end{equation}
Note that the projectors satisfy the following relations 
\begin{eqnarray}
    \vec{P}_\mathcal{B}^T\cdot \vec{P}_\mathcal{B} &=& 
    \vec{P}_\mathcal{A}^T\cdot \vec{P}_\mathcal{A} \ \ =  \ \ \bm{1}_{n_0 \times n_0}, \\
    \vec{P}_\mathcal{B}^T\cdot \vec{P}_\mathcal{X} &=& 
    \vec{P}_\mathcal{A}^T\cdot \vec{P}_\mathcal{X} \ \ =  \ \ \bm{0}_{n_0 \times (N-2n_0)},\\
    \vec{P}_\mathcal{X}^T\cdot \vec{P}_\mathcal{X}  &=&  \bm{1}_{(N-2n_0) \times (N-2n_0)}, \\
    \vec{P}_\mathcal{A}^T\cdot \vec{P}_\mathcal{B}&=&\bm{0}_{n_0 \times n_0}.
\end{eqnarray}

Defining the vectors $\Phi_{\pm1} = (\mathcal{A}_{\pm 1},\mathcal{B}_{\pm 1})^T$, the second step is to write the eigenvalue equation for the cell of the bend as
\begin{equation}\label{eq:EV_n0}
    \vec{P}_\mathcal{A} \vec{P}_\mathcal{B}^T \Phi_{-1} + \vec{H}_0^{\rm{bend}} \Phi_0 + \vec{P}_\mathcal{B} \vec{P}_\mathcal{A}^T\Phi_1 = E \Phi_0.
\end{equation}
$\bm{H}_0^{\rm{bend}}$ corresponds to the 
Hamiltonian of the isolated bend cell, i.e. the cell $n=0$ disconnected from the other supercells.
Using the expansion of $\Phi_0$ in terms of the projector operators, the above equation can be simplified to 
\begin{equation}\label{eq:projected_n0}
    \vec{P}_\mathcal{A}\mathcal{B}_{-1} + \vec{P}_\mathcal{B}\mathcal{A}_{1} = \left(E - \vec{H}_0^b\right) \left( \vec{P}_\mathcal{A} \mathcal{A}_0+ \vec{P}_\mathcal{B} \mathcal{B}_0  +\vec{P}_\mathcal{X} \mathcal{X}_0 \right). 
\end{equation}
We now introduce the Green function of the bend cell $\bm{\mathcal{G}}^{\rm{bend}} (E) = \left( E - \vec{H}_0^{\rm{bend}} \right)^{-1}$ written in block components
\begin{equation}
    \bm{\mathcal{G}}^{\rm{bend}} (E) = \begin{pmatrix}
        \bm{\mathcal{G}}^{\rm{bend}}_{aa} & \bm{\mathcal{G}}^{\rm{bend}}_{ab}& \bm{\mathcal{G}}^{\rm{bend}}_{ax} \\
        \bm{\mathcal{G}}^{\rm{bend}}_{ba} & \bm{\mathcal{G}}^{\rm{bend}}_{bb}& \bm{\mathcal{G}}^{\rm{bend}}_{bx} \\
        \bm{\mathcal{G}}^{\rm{bend}}_{xa} & \bm{\mathcal{G}}^{\rm{bend}}_{xb}& \bm{\mathcal{G}}^{\rm{bend}}_{xx}
    \end{pmatrix}.
\end{equation}
Explicitly, the block components can be obtained using the projector operators as follow
\begin{equation}
    \bm{\mathcal{G}}^{\rm{bend}}_{\alpha\beta}(E) = \vec{P}_\alpha^T \bm{\mathcal{G}}^{\rm{bend}}(E) \vec{P}_\beta,\quad \text{with} \quad \alpha,\ \beta \in \{ \mathcal{A}, \mathcal{B}, \mathcal{X}\}.
\end{equation}
Multiplying Eq.~\eqref{eq:projected_n0} from the left by $\vec{P}_\mathcal{A}$ and $\vec{P}_\mathcal{B}$, the eigenvalues equations can be written as
\begin{eqnarray}
    \mathcal{A}_0 &=& \bm{\mathcal{G}}^{\rm{bend}}_{ab} \mathcal{A}_1 + \bm{\mathcal{G}}^{\rm{bend}}_{aa} \mathcal{B}_{-1}, \label{eq:connection1}\\
    \mathcal{B}_0 &=& \bm{\mathcal{G}}^{\rm{bend}}_{bb} \mathcal{A}_1 + \bm{\mathcal{G}}^{\rm{bend}}_{ba} \mathcal{B}_{-1}, \label{eq:connection2}\\
    \mathcal{X}_0 &=& \bm{\mathcal{G}}^{\rm{bend}}_{xb} \mathcal{A}_1 + \bm{\mathcal{G}}^{\rm{bend}}_{xa} \mathcal{B}_{-1}.\label{eq:connection3}
\end{eqnarray}
Note that the amplitude $\mathcal{X}_0$ is entirely obtained from $\mathcal{A}_{1}$ and $\mathcal{B}_{-1}$. Hence one does not require Eq.~\eqref{eq:connection3} to relate the left and right solutions but will be used to reconstruct the solution inside the bend.

\subsubsection{Reflection and transmission across a bend}
Using the modal decomposition constructed from the transfer matrix of the ribbons, $\Psi_n$ outside the bend cell can be decomposed as a linear superposition of the ribbons eigenmodes.
We can then define a scattering problem by imposing specific boundary conditions.
A general scattering solution can be written as 
\begin{eqnarray}
    \Psi_n &=& 
    \sum_{j=1}^{n_0} a_j (\lambda_{L,j}^+)^n \psi_{L,j}^+ + r_j (\lambda_{L,j}^-)^n \psi_{L,j}^- \quad \text{for} \quad n\leq 0 \label{eq:scat1}\\
   &=& \sum_{j=1}^{n_0} t_j (\lambda_{R,j}^+)^n \psi_{R,j}^+ \quad \text{for} \quad n> 0.\label{eq:scat2}
\end{eqnarray}
The subscript $L$ indicates that these modes correspond to the ribbon on the left side of the bend, similarly, we will use the subscript $R$ to refer to modes in the ribbon on the right side of the bend.
The complex numbers $a_j$ , $r_j$ and $t_j$ represent the
amplitude of the incoming, reflected and transmitted modes in the scattering configuration. 
In particular, we defined the reflection and transmission matrices, respectively $\bm{R}$ and $\bm{T}$, via
\begin{equation}
    \hat{r} = \bm{R}\hat{a} \quad \text{and} \quad \hat{t} = \bm{T}\hat{a},
    \end{equation}
where 
    $\hat{a} =(a_1,\dots,a_{n_0})^T$, $\hat{r} =(r_1,\dots,r_{n_0})^T$ and $\hat{t} =(t_1,\dots,t_{n_0})^T$.
Comparing Eqs.~\eqref{eq:connection1} \&~\eqref{eq:connection2} with Eqs.~\eqref{eq:scat1} \&~\eqref{eq:scat2}, we can write the following system of equations satisfied by the reflection and transmission matrices
\begin{equation}
    \begin{pmatrix}
        -\bm{\mathcal{G}}_{ab}^b \bm{B}^-_L & \bm{B}^+_R \bm{D}^+_R - \bm{\mathcal{G}}_{aa}^b \bm{A}_R^+ \bm{D}_R^+ \\
        \bm{A}_L^- -\bm{\mathcal{G}}_{bb}^b \bm{B}_L^- & -\bm{\mathcal{G}}^b_{ba}\bm{A}_R^+ \bm{D}_R^+ 
    \end{pmatrix} \binom{\bm{R}}{\bm{T}} = \begin{pmatrix}
    \bm{\mathcal{G}}_{ab}^b \bm{B}_L^+ \\
        \bm{\mathcal{G}}_{bb}^b\bm{B}_L^+ - \bm{A}_L^+
    \end{pmatrix},
\end{equation}
which we invert to obtain $\bm{R}$ and $\bm{T}$.

As previously stated, we are focusing on an energy range where only a pair of propagating modes is present in the ribbon to the left of the bend. 
We assume that this pair of modes is the first pair of modes constructed previously, that is $(\psi_{L,j=1}^+,\psi_{L,j=1}^-)$. 
We therefore consider scattering solutions where a unique incoming mode of unit amplitude is impinging on the turn from the left which can be written as
\begin{eqnarray}
    \Psi_n &=& 
    (\lambda_{L,1}^+)^n \psi_{L,1}^+ +
    \sum_{j=2}^{n_0}  r_j (\lambda_{L,j}^-)^n \psi_{L,j}^- \quad \text{for} \quad n\leq 0 \label{eq:scat1}\\
   &=& \sum_{j=1}^{n_0} t_j (\lambda_{R,j}^+)^n \psi_{R,j}^+ \quad \text{for} \quad n> 0.\label{eq:scat2}
\end{eqnarray}

Since there exists a unique pair of propagating modes, we further define the reflection and transmission coefficients, respectively $R$ and $T$, for the propagating modes as 
\begin{equation}
    R = r_1= \bm{R}_{1,1} \quad \text{and} \quad T = t_1 = \bm{T}_{1,1}. 
\end{equation}
In addition to the reflected and transmitted wave, an incoming mode will excite left moving evanescent modes in the left ribbon and right moving evanescent modes in the right ribbon, i.e. $r_{j\neq1} \neq 0 $ and $t_{j\neq1} \neq 0 $.

Due to the mode normalisation, discussed in Sec.~\ref{subsec:properties}, the reflection and transmission coefficients satisfy the energy conservation condition
\begin{eqnarray}
    |R|^2 + |T|^2 = 1.
\end{eqnarray}

\section{Analysis}\label{sec:analysis}

\subsection{Geometrical configurations}

\begin{figure}
    \centering
    \includegraphics[scale = 0.8]{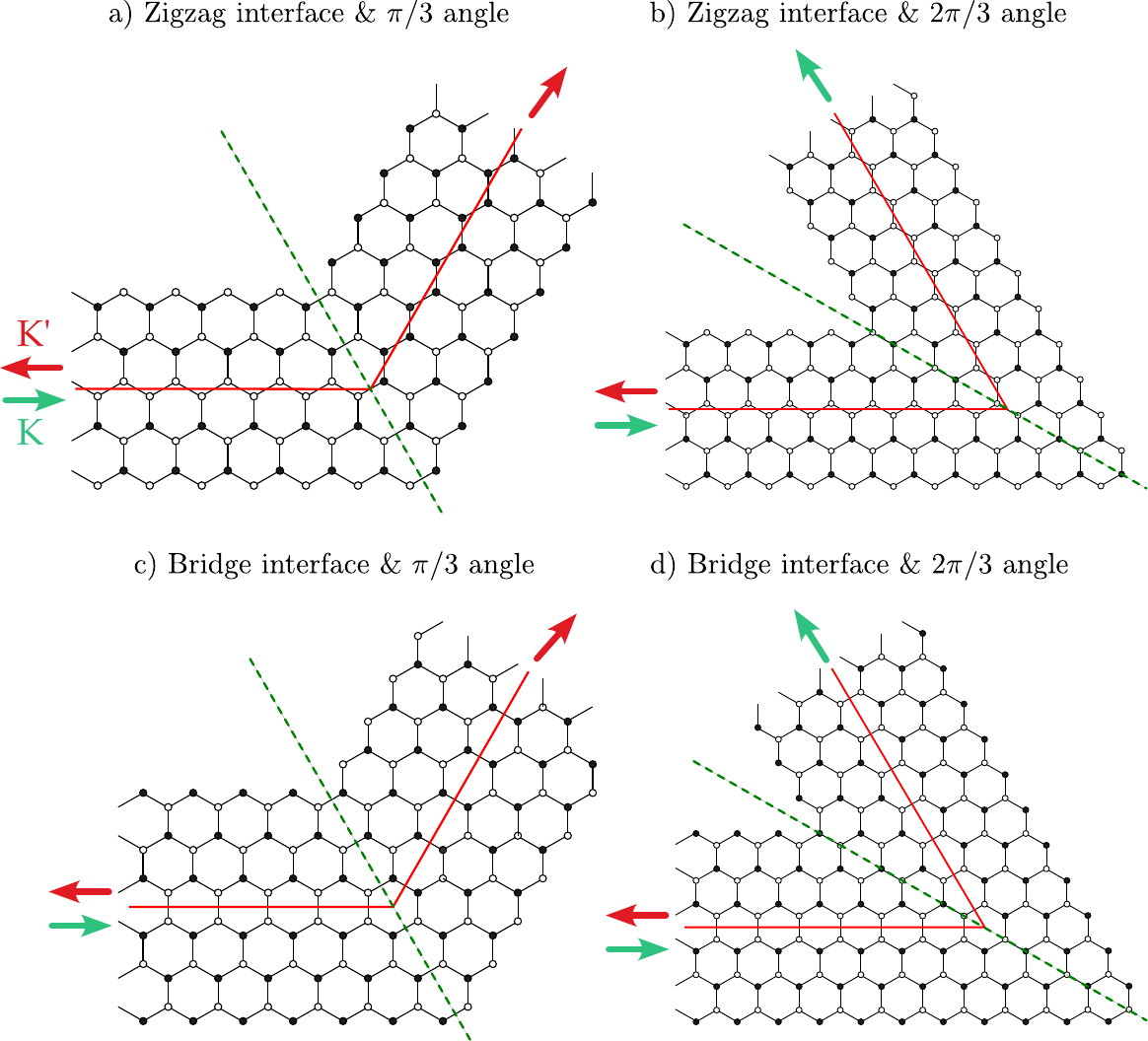}
    \caption{Different geometrical configurations combining the two possible interfaces and the two possible angles for the bend. The solid red line represents the interface: zigzag in Panels a) and b) and bridge in Panels c) and d). The green dashed line represents the axis involved in the reflection symmetries of the various configurations. The arrows represent the direction of propagation of the interface modes and their color indicates the valley index (see Fig.~\ref{fig:graphene_nogap_summary}).}
    \label{fig:geometrical_config}
\end{figure}

The two possible interfaces we are considering can be oriented along three possible directions in the graphene lattice, with only two (non trivial) angles.
Hence, the configurations we are interested in admit two natural angles as well as two possible interfaces, which constitute four canonical scattering configurations, represented in Fig.~\ref{fig:geometrical_config}.
In the following we will analyse the four configurations simultaneously and results associated with different configurations will be presented in the same format as in Fig.~\ref{fig:geometrical_config}.

Before discussing the scattering properties of the various configurations, we begin by noticing general properties of the different scenarios.
\begin{itemize}
    \item The $2\pi/3$ angle configurations exhibit a mirror symmetry with respect to the green dashed line in Figs.~\ref{fig:geometrical_config}-b) and d).
    \item The $\pi/3$ angle configurations are symmetric under the operation $\mathcal{M}\Gamma$, where $\mathcal{M}$ is the operator corresponding to mirror symmetry with respect to the green dashed line in Figs.~\ref{fig:geometrical_config}-a) and c), and $\Gamma$ is the chiral operator of Eq.~\eqref{eq:Gamma}. Indeed, we notice that the mirror operation flips the role of A and B, which means a flip of sign of the on-site potential. Subsequently applying $\Gamma$ flips back the on-site potential, so the overall configuration is invariant under the action of $\mathcal{M}\Gamma$.
    \item The onsite potential at the interface changes sign on each side of the bend in the case of a $\pi/3$ bend while it remains unchanged in the case of a $2\pi/3$ bend. This implies in particular that the scattering of a valley Hall mode in this configuration will necessarily flip its valley index since the reflected and transmitted waves lie in the opposite valley, as represented by the arrows in Fig.~\ref{fig:geometrical_config}.
\end{itemize}

These properties will have an impact on the transmission properties as we will see below.

\subsection{Example behaviour at fixed onsite potential and ribbon width}

In all canonical configurations, there are two external parameters that one may vary: the onsite potential $u$ and the ribbon width $N$.
We will study the dependence on each parameter in detail in the next subsections. 
Before entering this discussion, we describe the energy dependence of the reflection and transmission coefficient by studying a particular example where the onsite potential and the ribbon width are fixed.
This will allow us to present the general behaviour of the transmission and reflection spectra as well as to note some qualitative features that we should explore in the following.

\subsubsection{Spectrum}
\begin{figure}
    \centering
    \includegraphics[scale = 0.85]{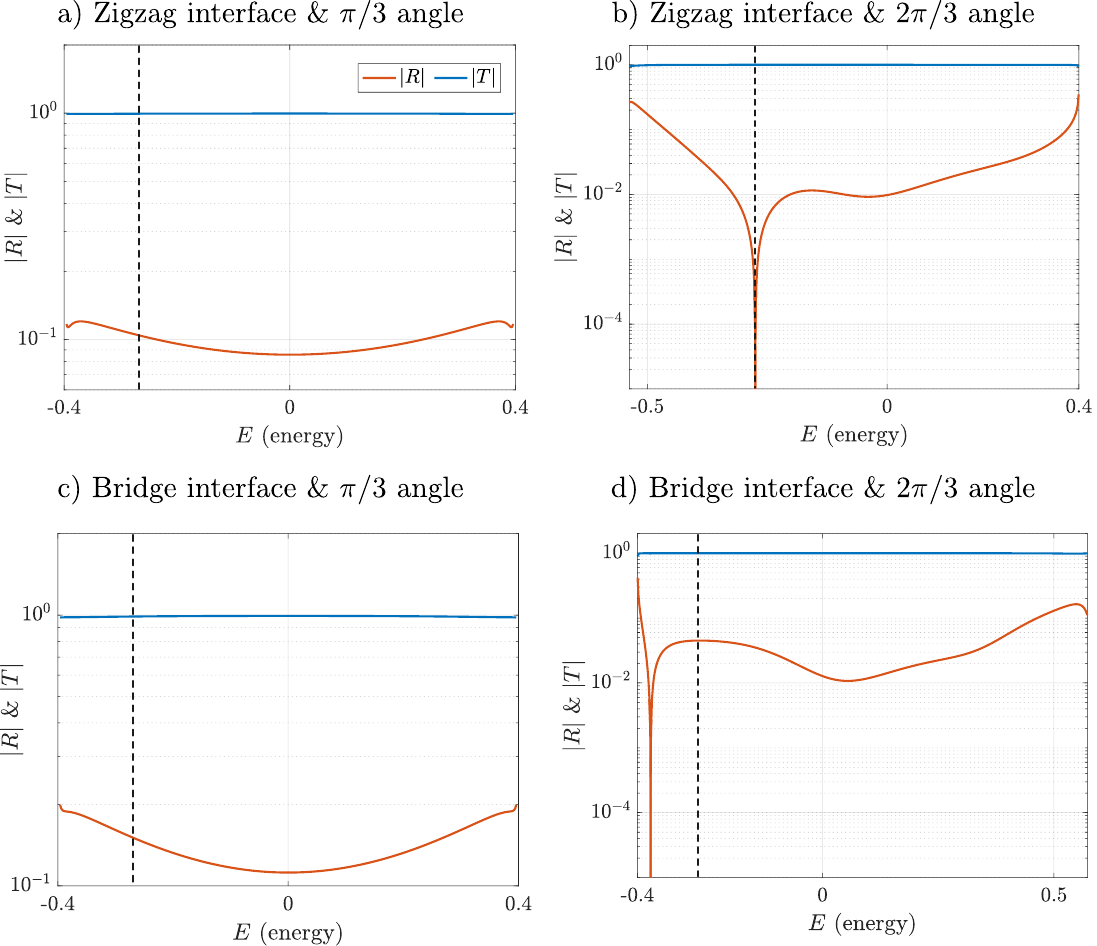}
    \caption{Scattering properties for a particular choice of $u$ and $N$ in all canonical geometries. In these examples, the onsite potential has been fixed to $u=0.4$ and the ribbon width is $N=12$ for the bridge interface and $N=13$ for the zigzag interface. The dashed vertical lines correspond to the energy \mbox{$E\approx-0.27$}. At this energy, the reflection coefficient vanishes for the case of a zigzag interface and a $2\pi/3$ angle of the bend (see Panel b)).}
    \label{fig:example}
\end{figure}

The reflection and transmission coefficients are computed over the energy range where there exists a single pair of propagating modes on each side of the bend and are presented in Fig.~\ref{fig:example}.
The main observation from this example is the fact that in all configurations, the reflection coefficient is relatively small and therefore the transmission is close to $1$.
This implies that, independently of the angle of the bend and of the interface type, valley-Hall modes are well transmitted across the bend.

There are two further observations that one can make.
First of all, it can been seen that the reflection coefficient vanishes for some values of the energy in the case of the $2\pi/3$ bend. The location of the zeros will be studied below and their presence can be linked to the mirror symmetry of this configuration.
Secondly, in the case of the $\pi/3$ bend, the spectrum is symmetric with respect to $E=0$. This is a consequence of the particular symmetry of this configuration.
Indeed, let us fix $E$ and $u$ and denote $\phi^{+}_{L(R)}(E,u)$ (resp. $\phi^{-}_{L(R)}(E,u)$) the right (resp. left) propagating modes in the ribbon on the left (right) to the bend.
We assume that we have a scattering solution, $\Phi(E,u)$ which takes the asymptotic form : 
\begin{eqnarray}
    \Psi_n(E,u) &\Equiv{n\rightarrow -\infty}& \psi^+_L(E,u) + R(E,u) \ \psi^-_L(E,u), \\
    &\Equiv{n\rightarrow +\infty}& T(E,u)\ \psi^+_R(E,u).
\end{eqnarray}
We recall from the last part of Sec.~\ref{sec:lattice} that the chiral operator, $\Gamma$, acts as 
\begin{equation}
    \Gamma\psi^\pm_{L/R}(E,u) =\psi^\pm_{L/R}(-E,-u),
\end{equation}
hence, we have that
\begin{eqnarray}
    \Gamma\Psi_n(E,u) &\Equiv{n\rightarrow -\infty}& \psi^+_L(-E,-u) + R(E,u) \ \psi^-_L(-E,-u), \\
    &\Equiv{n\rightarrow +\infty}& T(E,u)\ \psi^+_R(-E,-u).
\end{eqnarray}
Applying the mirror operator, $M$, which acts on the eigenmodes as
\begin{equation}
    \psi^\pm_{L/R}(-E,-u) = \psi^\pm_{R/L}(-E,u),
\end{equation} we get
\begin{eqnarray}
    M\Gamma\Psi_n(E,u) &\Equiv{n\rightarrow -\infty}& T(E,u)\ \psi^-_L(-E,u) \\
    &\Equiv{n\rightarrow +\infty}& \phi^-_R(-E,u) + R(E,u) \ \psi^+_R(-E,u).
\end{eqnarray}
Hence from the unicity of the scattering solution, $M\Gamma\Psi_n(E,u)$ is a solution to the scattering at energy $-E$, with reflection and transmission coefficients $R(-E,u)$ and $T(-E,u)$.
We can therefore read the following relations:
\begin{eqnarray}
    R(E) &=& R(-E) \\
    T(E) &=& T(-E),
\end{eqnarray}
which proves that the reflection and transmission coefficients are even functions of the energy.
This implies in particular that the point $E=0$ corresponds to a local extremum of the spectrum in the $\pi/3$ angle configurations.

\subsubsection{Visualisation of scattering solutions}

\begin{figure}[h!]
    \centering
\includegraphics[scale=0.7]{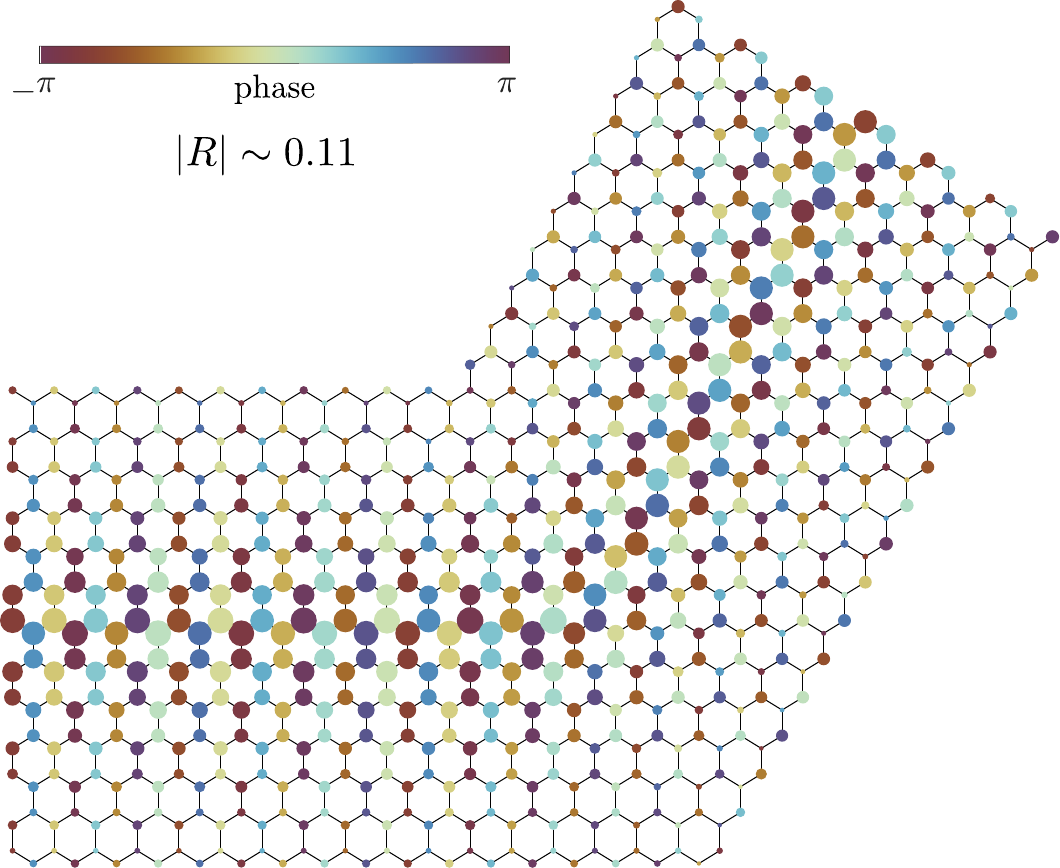}
    \caption{Mode reconstruction for the case of a zigzag interface and a bend with angle $\pi/3$ at a fixed energy \mbox{$E\approx-0.27$}. The ribbon width is chosen to be $N=13$ and $u=0.4$. The reflection coefficient for this configuration is $|R| \approx 0.11$.}
    \label{fig:particular_bad_zigzag}
\end{figure}

Using Eqs.~\eqref{eq:scat1} \&~\eqref{eq:scat2} as well as the reflection an transmission matrices, we can reconstruct the field in the ribbons of both sides of the bend as well as within the bend.
This allows to visualise the scattering solutions which we present below for a fixed value of the energy for all possible configurations.
The value of the energy is chosen to be $E\approx -0.27$ which corresponds to a zero of the reflection coefficient in the case of a zigzag interface and for the $2\pi/3$ bend (see Fig.~\ref{fig:example}-b) ).
\bigskip
\\
\textbf{Zigzag interface and $\pi/3$ bend:}
\\
Fig.~\ref{fig:particular_bad_zigzag} shows the reconstructed scattering solution for a zigzag interface and a bend with an angle of $\pi/3$. In this configuration the reflection coefficient is approximately $|R| \approx 0.11$.
Note that the transition from anti-symmetric to symmetric mode is not clearly visible, due to the geometrical configuration of the zigzag interface. Indeed the graphene ribbon with zigzag interface does not satisfy the mirror symmetry\footnote{
However, the associated Rice-Mele chain obeys the mirror symmetry. Hence it is the eigenmodes of the Rice-Mele and not directly of the graphene ribbon which display symmetric and anti-symmetric behaviours.}.
\\
\textbf{Zigzag interface and $2\pi/3$ bend:}
\begin{figure}[h!]
    \centering
\includegraphics[scale=0.7]{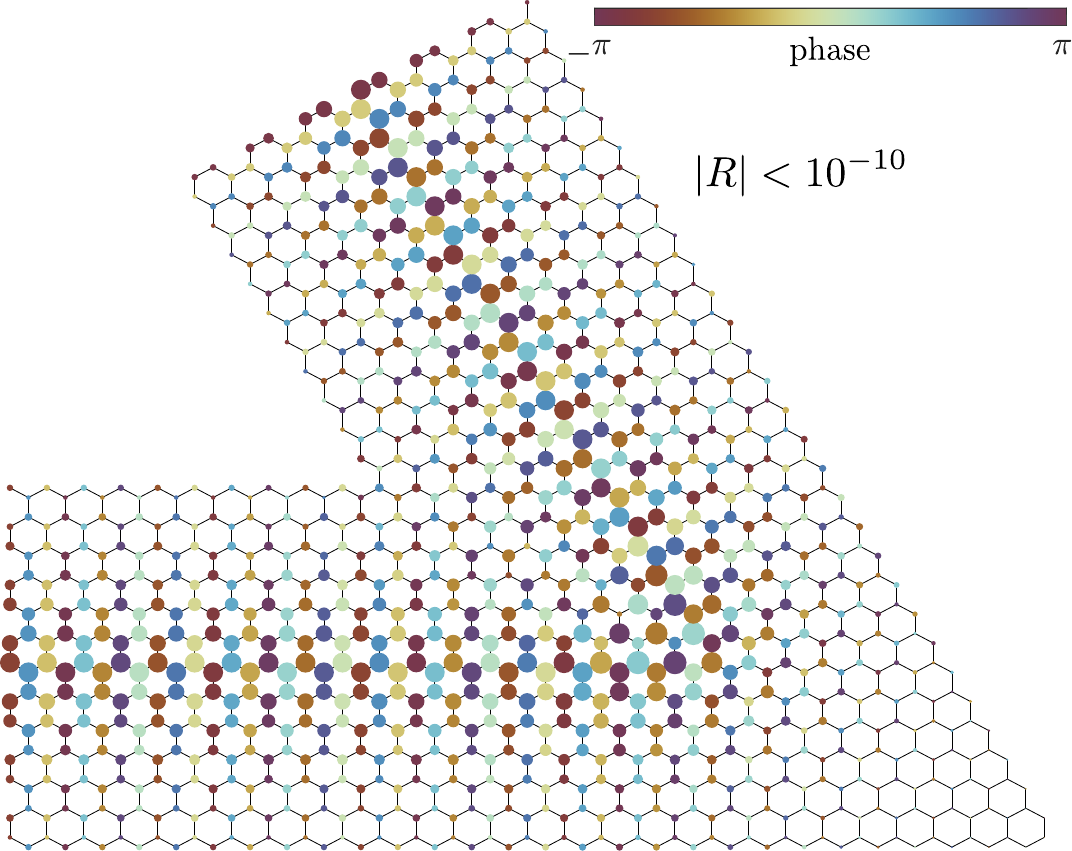}
    \caption{Mode reconstruction for the case of a zigzag interface and a bend with angle $2\pi/3$ at a fixed energy \mbox{$E\approx-0.27$}. The ribbon width is chosen to be $N=13$ and $u=0.4$. The reflection coefficient for this configuration is $|R| < 10^{-10}$, which corresponds to a perfect transmission of the interface mode.}
    \label{fig:particular_good_zigzag}
\end{figure}
\\
Fig.~\ref{fig:particular_good_zigzag} shows the reconstructed scattering solution for a zigzag interface and a bend with an angle of $2\pi/3$. For this configuration, the absolute value of the reflection coefficient is less than $10^{-10}$, which indicates a perfect transmission of the mode through the bend. 
Note that visually, the modes in the ribbon to the left of the bend appear similar in Fig.~\ref{fig:particular_bad_zigzag} and Fig.~\ref{fig:particular_good_zigzag}. However in the latter case, backscattering is completely suppressed ($|R|<10^{-10}$) while it is non-negligible in the former ($|R|\approx 0.11$).
Hence, in order to argue the immunity to backscattering of valley state, one has to perform a detailed quantitative analysis.
\\
\textbf{Bridge interface and $\pi/3$ bend:}
\begin{figure}
    \centering
\includegraphics[scale=0.7]{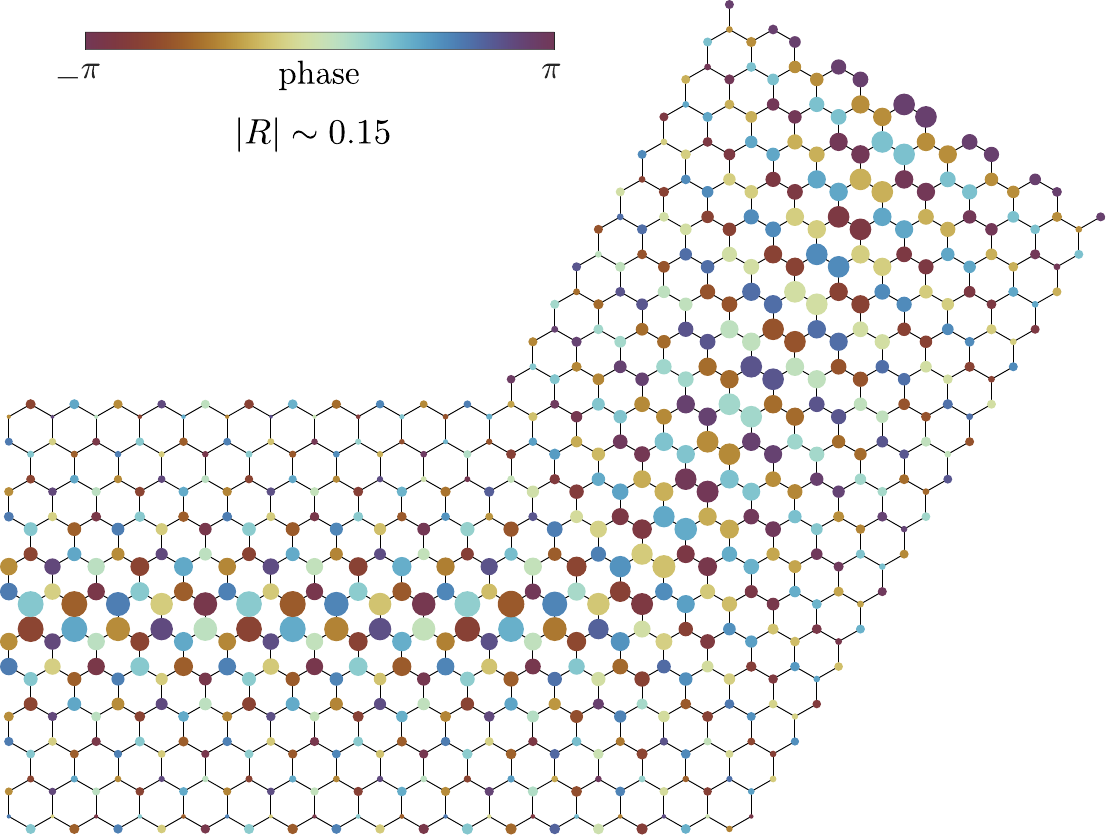}
    \caption{Mode reconstruction for the cas of bridge interface and a bend with angle $\pi/3$ at a fixed energy \mbox{$E\approx-0.27$} (see Fig.~\ref{fig:example}. The ribbon width is chosen to be $N=12$ and $u=0.4$. The reflection coefficient for this configuration is \mbox{$|R| \approx 0.15$}.}
    \label{fig:particular_bad_bridge}
\end{figure}
\\
Fig.~\ref{fig:particular_bad_bridge} shows the reconstructed scattering solution for a bridge interface and a bend with an angle of $\pi/3$. 
Note that the sites closest to the interface on the ribbon to the left of the bend have a positive onsite potential while the ones on the right of the bend have a negative onsite.
This implies in particular that the interface states are anti-symmetric in the left ribbon and symmetric in the right ribbon, as shown in Sec.~\ref{App:RM} of the supplemental material.
This transition between anti-symmetric to symmetric modes is clearly visible in the reconstruction.
\\
\textbf{Bridge interface and $2\pi/3$ bend:}
\begin{figure}
    \centering
\includegraphics[scale=0.7]{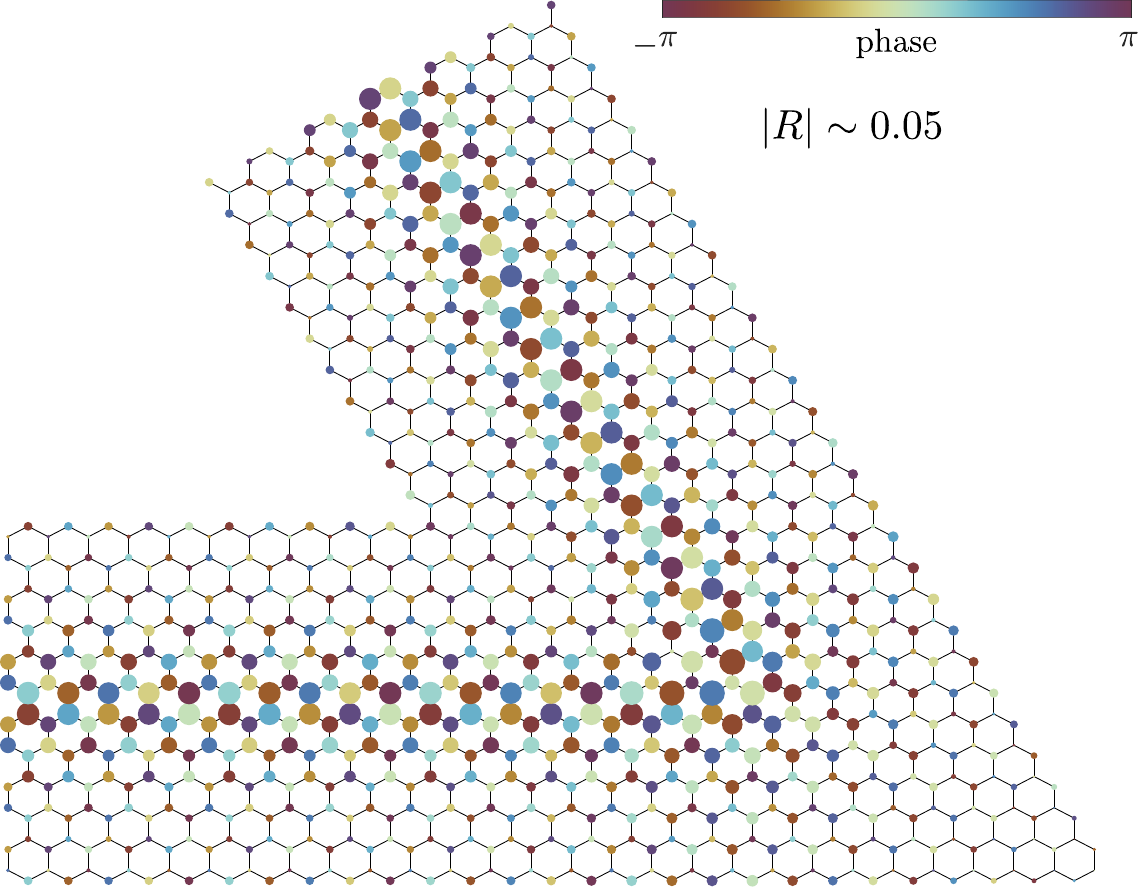}
    \caption{Mode reconstruction for the case of a bridge interface and a bend with angle $2\pi/3$ at a fixed energy $E\approx-0.27$. The ribbon width is chosen to be $N=12$ and $u=0.4$. The reflection coefficient for this configuration is \mbox{$|R| \approx 0.05$.}}
    \label{fig:particular_good_bridge}
\end{figure}
\\
Fig.~\ref{fig:particular_good_bridge} shows the reconstructed scattering solution for a bridge interface and a bend with an angle of $2\pi/3$.
In this geometrical configuration, the ribbons on each sides of the bend have the same structure and therefore the modes are identical on both sides of the bend.
We can see that the anti-symmetric interface state impinging from the left is well transmitted to the same anti-symmetric interface state in the ribbon on the right of the bend.
For these parameters, the amount of reflection is approximately $0.05$.

\subsection{Influence of the ribbon width}

After having discussed the general scattering properties in a particular example we now turn our attention to the influence on the parameters, beginning with the impact of the ribbon width $N$. In this section, we fix the value of the onsite potential to be $u=0.3$.

\begin{figure}[h]
    \centering
    \includegraphics[scale = 0.8]{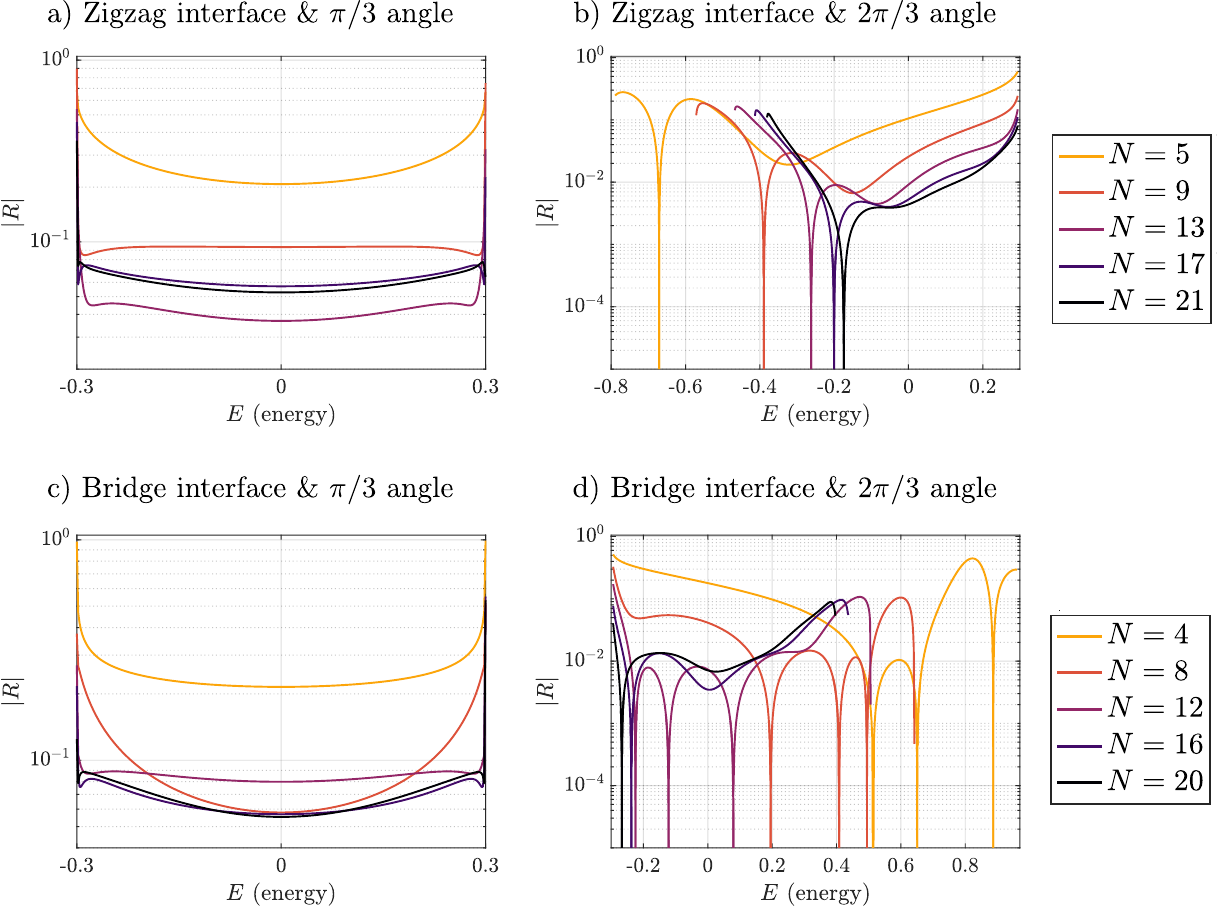}
    \caption{Influence of the ribbon width $N$ on the reflection coefficient for fixed onsite potential $u=0.3$. For this value of the onsite potential, we have $N_c\approx 3.4$.}
    \label{fig:influence_N}
\end{figure}

Fig.~\ref{fig:influence_N} depicts the influence of the ribbon width on the reflection spectrum in all canonical configurations. We have plotted the reflection coefficient for various values of $N$ ranging from $N=4\ (+1)$ to $N=20\ (+1)$ in the case of the bridge (zigzag) interface.

The first observation valid for all configurations is that overall, the reflection decreases over the entire energy range as $N$ increases.

In the case of the $\pi/3$ bend and $u=0.3$, the curves converge to a relatively flat spectrum at the value $|R|\approx 0.06$ for both the zigzag and the bridge interfaces.

In the $2\pi/3$ case, the two interfaces present distinct behaviours. In particular the bridge interface configuration seems to exhibit more zeros than the zigzag interface case. The location of these zeros is influenced by the ribbon width.
For this angle however, it appears that the reflection coefficient converges to a curve with a single zero and with similar qualitative behaviour for the two interfaces. 

\subsection{Influence of the onsite potential}

\begin{figure}[h]
    \centering
    \includegraphics[scale = 0.8]{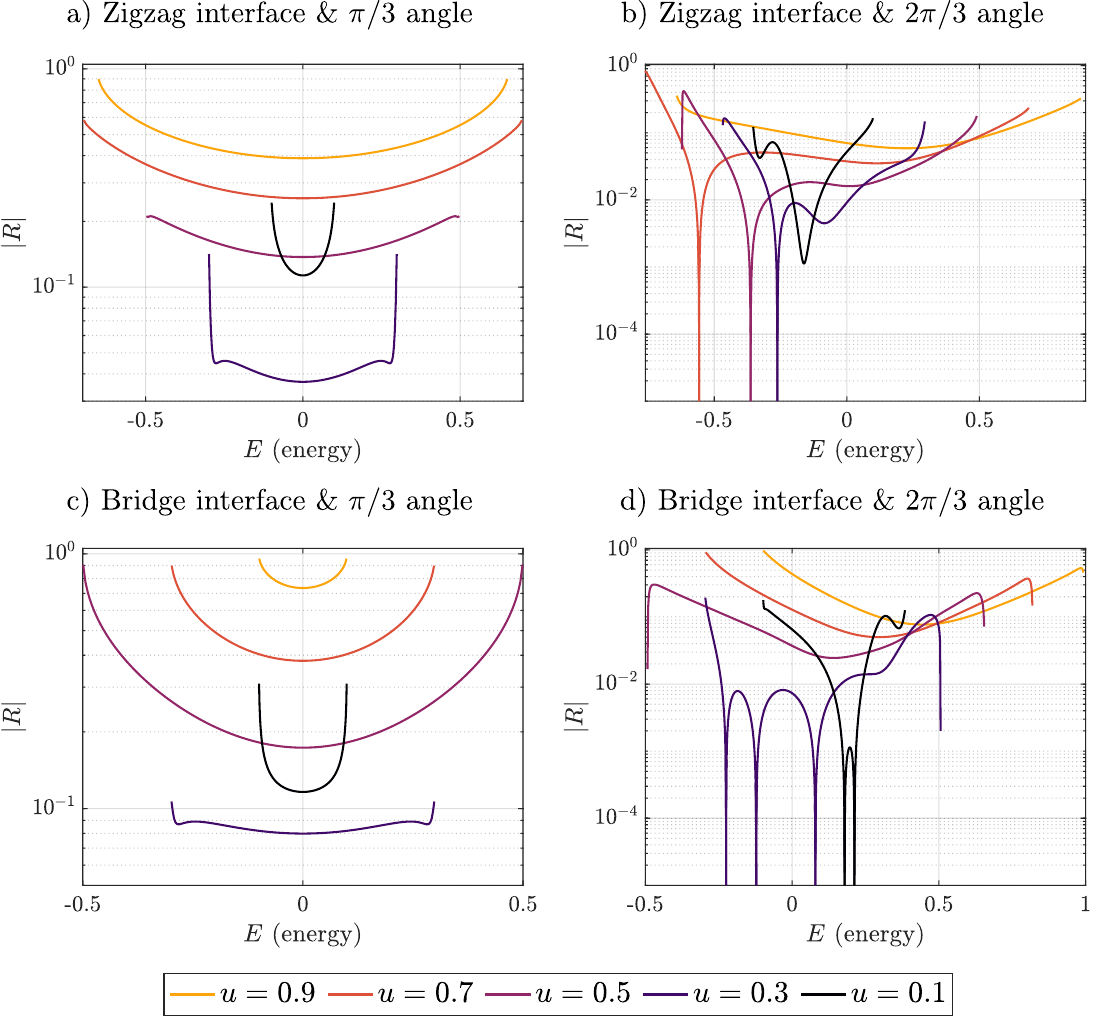}
    \caption{Influence of the onsite potential $u$ on the reflection coefficient. The width of the ribbon is fixed to $N=12$ for the bridge interface and $N=13$  for the zigzag interface. For these ribbon widths, we have $u_c \approx 0.08$.}
    \label{fig:influence_u}
\end{figure}

Fig.~\ref{fig:influence_u} depicts the influence of changing the value of the onsite potential on the different bends and the different interfaces.
As in the case of the ribbon width, the onsite potential has a non-trivial influence on the reflection coefficient in both configurations.
In the case of the $\pi/3$ bend, decreasing the value of the onsite potential seems to reduce the amount of reflection up to a certain value where the reflection increases as $u$ decreases. 
This therefore suggests that there exists an optimum value of $u$ which would minimise the reflection for the $\pi/3$ bend configuration. 

A case can be made for a similar statement in the bend with an angle of $2\pi/3$, althought the reflection coefficient fluctuates more than in the $\pi/3$ case. 
Indeed, it seems that decreasing the value of $u$ up to some value does decrease the overall reflection coefficient across the energy range available.
The value of the onsite potential does also influence the location and number of zeros of the reflection coefficients.

Overall, we note that the dependence of the reflection coefficient on the external parameters, $u$ and $N$, does not display a monotonous nor simple behavior. In particular, even though the modes are localised near the interface, the localisation length depends on those external parameters. To estimate this, a site localisation number $N_{\rm loc}= - \ln(\lambda_-)^{-1}$ can be defined, where $\lambda_-$ is the eigenvalue given in Eq.~\eqref{eq:EV_lambda}. When the width of the ribbon is comparable to $N_{\rm loc}$, finite size effects are expected to play a significant role. Using the results detailed in Appendix~\ref{App:RM}, we have that $N_{\rm loc} \geq N_c=\ln(u - \sqrt{1+u^2})^{-1}$, which implies that $N_{\rm loc}$ goes to infinity when $u$ goes to~0. 
Equivalently, we can define a critical onsite potential $u_c$ as a function of $N$ by inverting the previous relation.
Since the mode profiles
are intrinsically different whether $N$ is larger or smaller than $N_c$ (equivalently with $u$ and $u_c$), it is natural to
expect different behaviours in the two regimes. For example, when $u=0.3$ as in Fig.~\ref{fig:influence_N}, we have $N_c \approx 3.4$.

\section{Conclusion}\label{sec:conclusion}

In this paper, we have quantitatively studied the propagation and scattering of valley Hall interface modes in discrete lattice models which emerge in asymptotic regimes of phononics and photonics crystals and of acoustic networks. In particular, we have focused on honeycomb lattices and their arrangement in ribbon configurations.

After reviewing the general construction of valley Hall modes in graphene lattices with the addition of an onsite potential, we have investigated their realisation in ribbons with zigzag edges.
We have analysed the dispersion relation of graphene ribbons with zigzag edges and recalled that the addition of an onsite potential will open a gap in its band structure. 
We have then showed that if one switches the sign of the onsite potential in part of the ribbon, that is if one creates an interface, one can create additional modes localised near the interface with energies inside the band gap.
Those modes are usually referred to as valley Hall modes.

From the description of valley Hall modes in ribbons we then described their propagation and scattering through canonical obstacles, namely sharp bends at angles of $\pi/3$ and $2\pi/3$. 
Such scattering scenarios are widely used in experimental and numerical studies to assess the robustness and backscattering immunity of valley Hall modes, some features oftentimes linked with the idea of topological protection of the valley index.

However, due to the presence of the time-reversal symmetry, the decoupling between the valley is approximate and one should always expect partial reflection in a general scattering situation.
We therefore quantitatively characterized the amount of transmission and reflection in four specific configurations corresponding to the two possible angles of the bend and the two possible interface types of ribbons with zigzag edges. 

To do so, we have defined reflection and transmission coefficients associated with propagating modes in the ribbons on each side of the bends. 
These quantities were constructed using the transfer matrix formalism and computed numerically up to machine precision. 
In addition to constructing scattering quantities far from the bend, this method allowed us to access the excitation of evanescent modes and to reconstruct and visualize the near-field scattering solutions in the vicinity of the bend.
An example of such transmission and reflection spectra was presented in Fig.~\ref{fig:example} and the associated solutions in Figs.~\ref{fig:particular_bad_zigzag},~\ref{fig:particular_good_zigzag},~\ref{fig:particular_bad_bridge} and~\ref{fig:particular_good_bridge}.
The main observations, in our model, can be summarized as follow:
\begin{itemize}
    \item The transmission of valley Hall modes through sharp bends is close to being maximal for all energies where only the interface modes are able to propagate.
    \item The reflection and transmission spectra are symetric with respect to $E=0$ for the bend at an angle of $\pi/3$. This is consequence of the combination of the chiral and mirror symmetry of the configuration.
    \item In the case of the $2\pi/3$ angle, we have shown the existence of energies for which the reflection is zero (up to numerical accuracy), leading to perfect transmission for finite and discrete values of $E$.
    \item The details of the reflection spectra change significantly with the external parameters of our model, namely the value of the onsite potential $u$ and the ribbon width $N$ (see Figs.~\ref{fig:influence_N} and~\ref{fig:influence_u}).
\end{itemize}

Our observations reinforce the idea that valley conservation is not correlated with a high transmission of interface modes across bends. In fact, in some configuration, high transmission occurs in spite of a change of the valley index. This is true for the case of a sharp turn at an angle $\pi/3$ as investigated in this study or in the case of a straight ribbon where the onsite potential is suddenly flipped, as shown in the supplementary material Sec.~\ref{sec:Shift}. 
Indeed, we have observed high transmission of interface modes across $\pi/3$ bends, a configuration which connects different valleys on each side of the defect. To further support this idea, we have compared in the supplemental material Sec.~\ref{Sec:Trivial}, the scattering coefficients of valley Hall modes to those of a bulk mode of a graphene ribbon, and it was observed that both the interface and the bulk modes have similar reflection and transmission coefficients in that specific geometrical configuration.

Finally, while the main part of the study was devoted to the propagation of valley Hall modes, we show in the supplementary material Sec.~\ref{Sec:armchair} that the tools and methods applied to ribbons with zigzag edges can be generalized to characterise the propagation in arbitrary ribbons. As an illustration, we have described and computed the reflection and transmission coefficients of interface modes in ribbons with armchair edges through a bend at a $\pi/3$ angle. We expect that our method can be used to further characterise mode propagation in a wide range of discrete lattices.
The investigation of scattering scenarios in tight-bindings models with different geometries through the efficient and accurate transfer matrix formalism may then guide the design of future continuous phononics and photonics crystals and other metamaterials.

A particularly interesting future application of our methods would be the quantitative study of backscattering in analogue QSH systems. As previously mentioned, analogue QSH devices strongly rely on spatial symmetries to allow for the existence of topological protected modes. These spatial symmetries are inevitably broken when considering finite size systems and defects. For this reason, and because analogue QSH systems are reciprocal, we do not expect their edge modes to be perfectly immune to backscattering. Applying our methods to QSH systems would allow for a better characterisation of the scattering properties of edge modes in such devices and in classical topological insulators more generally.

\vskip6pt

\enlargethispage{20pt}
\section*{Acknowledgement}This work received support from the French government under the France
2030 investment plan, as part of the Initiative d’Excellence
d’Aix-Marseille Université - A*MIDEX - AMX-19-IET- 010.


\vskip2pc

\bibliographystyle{RS}
\bibliography{sample}

\newpage
\appendix
\section{Berry curvature of a two level system}\label{App:Berry}
Here we present the details to compute the Berry curvature of the graphene lattice with onsite potential~\cite{asboth2016short}.
The Bloch Hamiltonian \eqref{HB_onsite} is a $2\times 2$ matrix which can be decomposed onto the Pauli matrices $\hat{\bm{\sigma}} = (\bm{\sigma}_x,\bm{\sigma}_y,\bm{\sigma}_z)$ (there is no contribution from $\bm{\sigma}_0$) as $\bm{H}_k = \hat{d}\cdot\hat{\bm{\sigma}}$. The components of $\hat{d}=(d_x,d_y,d_z)$ are given by
\begin{eqnarray}
    d_x &=& \mathrm{Re}[f(k)], \\
    d_y &=& \mathrm{Im}[f(k)], \\
    d_z &=& u.
\end{eqnarray}
With this notation, the two energy bands are given by $E_\pm = \pm |\hat{d}|$.
To find the eigenvectors associated to the two energy bands, it is convenient to introduces the unit vector $\tilde{d} = \hat{d}/|\hat{d}|$.
Since $\bm{H}_k = |\hat{d}|\tilde{d}.\bm{\sigma}$, finding eigenvectors associated with $E_\pm$ is equivalent to finding eigenvectors of the operator $\tilde{d}.\bm{\sigma}$ associated with eigenvalues $\pm 1$.
Those are easily expressed in terms of the polar decomposition $(\theta_k,\varphi_k)$ of $\tilde{d}$
\begin{equation}
    \tilde{d} = \begin{pmatrix}
        \sin{\theta_k} \cos{\varphi_k} \\
        \sin{\theta_k} \sin{\varphi_k} \\
        \cos{\theta_k}
    \end{pmatrix} \quad \text{with} \quad \theta_k = \arccos{\frac{d_z}{|\hat{d}|}} \quad \text{and} \quad \varphi_k = \arctan{\frac{d_x}{d_y}}.
\end{equation}
The eigenvectors $(u_-,u_+)$ are then given by
\begin{equation}
  u_- = \binom{\sin{\frac{\theta_k}{2}}}{-e^{i\varphi_k}\cos{\frac{\theta_k}{2}}}  \quad \text{and} \quad u_+ = \binom{\cos{\frac{\theta_k}{2}}}{e^{i\varphi_k}\sin{\frac{\theta_k}{2}}} .
\end{equation}
The Berry connection $\vec{\mathcal{A}}_\pm$ of the positive (negative) band is given by
\begin{equation}
    \vec{\mathcal{A}}_\pm = i \langle u_\pm|\nabla_{\vec{k}}|u_\pm \rangle.
\end{equation}
Explicitely, we have
\begin{eqnarray}
   \vec{\mathcal{A}}_+ &=& -\sin^2\left[\frac{\theta_k}{2}\right]\nabla_{\vec{k}} \varphi_k \quad \text{and}\\
\vec{\mathcal{A}}_- &=& -\cos^2\left[\frac{\theta_k}{2}\right]\nabla_{\vec{k}} \varphi_k.
\end{eqnarray}
The two are related via $\vec{\mathcal{A}}_+ + \vec{\mathcal{A}}_- = -\nabla_{\vec{k}} \varphi_k$. 
The Berry curvature $\mathcal{F}$ is then obtained as the curl of the Berry connection
\begin{eqnarray}
    \mathcal{F} &=& \nabla_{\vec{k}} \times \vec{\mathcal{A}}_\pm \\
    &=& \frac{1}{2} \sin{\theta} \left( \frac{\partial \theta_k}{\partial k_x}\frac{\partial\varphi_k}{\partial k_y} - \frac{\partial\varphi_k}{\partial k_x}\frac{\partial \theta_k}{\partial k_y} \right).
\end{eqnarray}
This is the quantity plotted in panel (b) of Fig.\ref{fig:graphene_nogap_summary}.

\section{Properties of the Rice-Mele chain}\label{App:RM}
We have seen in the main part of the text that the spectrum of graphene ribbons with an onsite potential is identical to the one of the Rice-Mele chain where the intracell hopping parameter is $k$ dependent and complex and the intercell hopping is equal to unity. Identically, the spectrum of ribbon with an interface can be described as the spectrum of two connected Rice-Mele chain with opposite onsite potentials in each chain.
In the following, we analytically investigate the properties of this spectrum in two limits : i) the fully dimerized limit and ii) the ribbon with infinite width.

\subsection{Spectrum in the fully dimerized limit}

\begin{figure}
    \centering
    \includegraphics[scale=0.8]{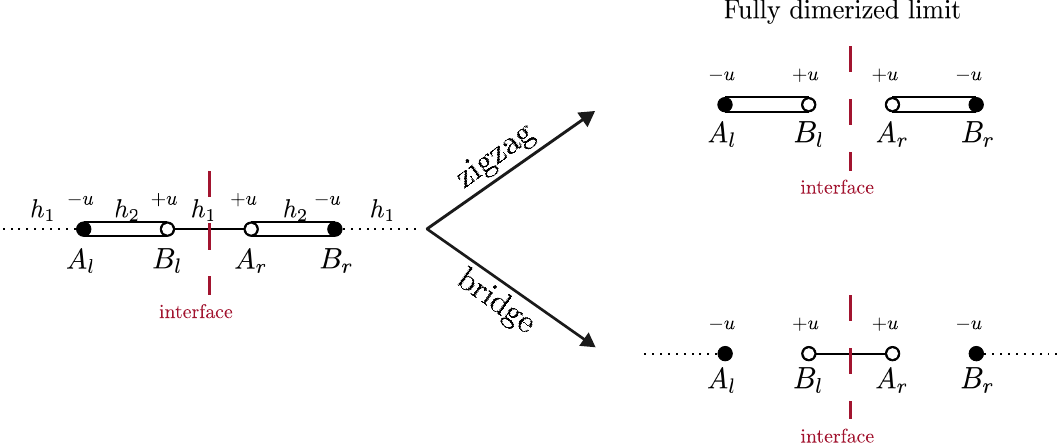}
    \caption{Schematic of the Rice-Mele chain with an interface located at an intercell hopping and its representation in the fully dimerized case for both interfaces.}
    \label{fig:RM_chain_dimerized}
\end{figure}

One can gain valuable insight by considering the fully dimerized limit, that is the case where the intracell hopping vanishes. 
We consider the general Rice-Mele chain with intracell hopping $h_1$, intercell hopping $h_2$ and onsite potential $u$. A schematic of the chain is shown in Fig.~\ref{fig:RM_chain_dimerized}.
To connect the Rice-Mele chain to the graphene ribbon, we identify $(h_1,h_2)$ with $(s(k),1)$ in the case of a zigzag interface and with $(1,s(k))$ in the case of a bridge interface.
Therefore the fully dimerized limit is obtain when $s(k) = 0$ that is at the edge of the Brillouin zone.

It can be seen in Fig.~\ref{fig:RM_chain_dimerized} that both interface has a very different behavior in the fully dimerized limit.
In the case of zigzag interface, sites on each site of the interface are not connected and are identical to other dimers in the each. Hence in this limit, interface states have the same energy spectrum that the rest of the bulk of the chain, that is $E=\pm \sqrt{1+u^2}$.
In the case of the bridge interface, a pair of site is connected by hopping unity and both have identical onsite potential.
The interface states are therefore the eigenstates of the Hamiltonian describing this dimmer, given by
\begin{equation}
    \bm{H}_d = \begin{pmatrix}
        u & 1 \\
        1 & u
    \end{pmatrix}.
\end{equation}
Its eigenvalues are $E=u\pm 1$ which are the energies of the interface states.

Regarding the edge states, for both interfaces in the fully dimerized limit, the edge of the chain are made of isolated sites with onsite potentials. Therefore, the energy of the edge states are equal to the value of the onsite potential on the site at the edge of the chain. Assuming that sites nearest to the interface have positive onsite potential, this implies that the edge modes have energy $E=u$ for a zigzag interface and $E=-u$ for a bridge interface.

\subsection{Interface states of the periodic chain}

\begin{figure}
    \centering
    \includegraphics[scale=0.9]{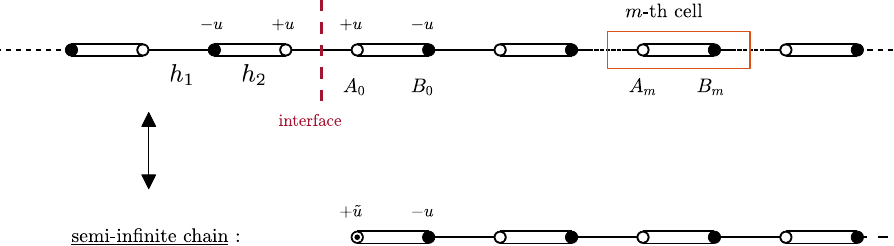}
    \caption{Schematic of the infinite Rice-Mele chain and its reduction to a semi-infinite chain.}
    \label{fig:schematic_RM_chain}
\end{figure}

Another instructive limit is the one of an infinite Rice-Mele chain, which corresponds to a ribbon with infinite width.
We derive here analytically the expression for the energy as well as for the eigenvectors of interface states in a ribbon with infinite transverse width. 
The Rice-Mele chain with an interface is shown in Fig.~\ref{fig:schematic_RM_chain}. Using the reflexion symmetry of the chain at the interface, we can look for symmetric and anti-symmetric modes and therefore look only at a semi-infinite chain depicted at the bottom of Fig.~\ref{fig:schematic_RM_chain}. 

Note that the first site of the semi-infinite chain will have a different onsite potential, $\tilde{u}$, which depends on the original on-site potential, the hopping across the interface as well as whether the mode is symmetric or anti-symmetric. To find the value of the on-site potential, we use the equation of motion on the site $A_0$:
\begin{equation}\label{eq:A0_eq}
    E A_0 = u A_0 + h_1 B_{-1}, + h_2 B_0.
\end{equation}

We now introduce the parameter $\sigma$, which allows us to distinguish between the symmetric ($\sigma = 1$) and anti-symmetric ($\sigma = -1$) modes. We therefore have $A_0 = \sigma B_{-1}$. Plugging this into the Eq.~\eqref{eq:A0_eq} gives:
\begin{equation}\label{eq:ic_RM}
    E A_0 = (u + h_1\sigma) A_0 + h_2 B_0 = \tilde{u} A_0 + h_2 B_0.
\end{equation}

To construct the interface solution of the infinite Rice-Mele chain, or equivalently the edge modes of the semi-infinite modified Rice-Mele chain, we first need to characterise the propagation in the bulk of the chain. 
The chain is made of unit cell composed of two sites, as shown in Fig.\ref{fig:schematic_RM_chain}. 
We now construct the transfer matrix from the $m$-th cell to the $m+1$-th cell. 
The equation of motion on the $B_m$ site gives:
\begin{equation}
    E B_m = -u B_m + h_2 A_m + h_1 A_{m+1},
\end{equation}
which can be rewritten, for $h_1\neq 0$ as
\begin{equation}\label{transfer_Am}
    A_{m+1} = \frac{E+u}{h_1} B_m - \frac{h_2}{h_1}A_m.
\end{equation}
The equation of motion on the $A_{m+1}$ site gives
\begin{equation}
    E A_{m+1} = uA_{m+1} + h_1B_m + h_2B_{m+1},
\end{equation}
which can be rewritten, for $h_2 \neq 0$ as
\begin{eqnarray}
    B_{m+1} &=& \frac{E-u}{h_2}A_{m+1} - \frac{h_1}{h_2}B_m \\
            &=& \frac{E^2 - u^2-h_1^2}{h_1h_2}B_m - \frac{E-u}{h_1}A_m,\label{transfer_Bm}
\end{eqnarray}
where we have used Eq.~\eqref{transfer_Am} in the second line. Combining Eqs.~\eqref{transfer_Am} and \eqref{transfer_Bm}, we can build the transfer matrix $\bm{M}$:
\begin{equation}
    \bm{M} = \begin{pmatrix}
-\frac{h_2}{h_1} & \frac{E+u}{h_1} \\
-\frac{E-u}{h_1}  & \frac{E^2 - u ^2 - h_1^2}{h_1h_2}
\end{pmatrix},
\end{equation}
which satisfies
\begin{equation}
    \begin{pmatrix}
A_{m+1} \\
B_{m+1}
\end{pmatrix} = \bm{M} \begin{pmatrix}
A_{m} \\
B_{m}
\end{pmatrix}.
\end{equation}
We can therefore construct the solution anywhere in the chain from the matrix $\bm{M}$ and the initial condition on $(A_0,B_0)$. Using Eq.\eqref{eq:ic_RM}, we have that $A_0 = h_2/(E-\tilde{u}) B_0$, and hence
\begin{equation}
    \begin{pmatrix}
A_{m} \\
B_{m}
\end{pmatrix} = \bm{M}^m \begin{pmatrix}
A_{0} \\
B_{0}
\end{pmatrix} = B_0 \bm{M}^m \begin{pmatrix}
\frac{h_2}{E-\tilde{u}} \\
1
\end{pmatrix}.
\end{equation}
The matrix $\bm{M}$ has 2 eigenvalues, $\lambda_{\pm}$ which satisfy $\lambda_+\lambda_{-} = 1$. 
Therefore, we either have \mbox{$\lambda_{-}<\lambda_+$} or \mbox{$\lambda_+ = \lambda_{-}^*$}. 
We are interested in the case where an interface is present which correspond to exponentially decaying modes along the chain. To each eigenvalue is associated an eigenvector $V_{\pm}$. An interface mode exists if the vector $(A_0,B_0)^T$ is aligned with $V_{-}$. Solving this condition leads to the following relation for $E$:
\begin{eqnarray}\label{eq:energy_interface}
    E_{i} = \sigma h_1 + \mathrm{sign}(u_{i}) \sqrt{h_2^2 + u^2}.
\end{eqnarray}
In the above formula, $u_{i}$ is the value of the onsite potential on the site nearest to the interface in the infinite chain.
Substituting Eq.\eqref{eq:energy_interface} into the eigenvalues of the matrix $\bm{M}$ we get the value for $\lambda_{-}$ for the symmetric and anti-symmetric modes:
\begin{equation}\label{eq:EV_lambda}
    \lambda_{-}(\sigma) = \sigma \Lambda = \sigma \left(\frac{u - \sqrt{h_2^2 + u^2}}{h_2}\right).
\end{equation}

The above formula indicates that in the case where the interface is across a hopping $h_1$ (which correspond to a zigzag interface of the graphene ribbon), the decay of the interface modes in the transverse direction \emph{do not depend} on $h_1$, or equivalently $k$. 
To those eigenvalues correspond the following eigenvectors
\begin{equation}\label{eq:eigenvec}
    V_{-}(\sigma) =  \begin{pmatrix}
\frac{h_2^2 + \sigma h_1 (u + \sqrt{h_2^2 + u^2})}{h_2 (\sigma h_1 - u + \sqrt{h_2^2 + u^2})} \\
1
\end{pmatrix}.
\end{equation}

We can connect Eq.~\eqref{eq:energy_interface} to the expressions obtained above in the fully dimerized case. 
In the case of a zigzag interface, we have $h_1 = 0$ and $h_2=1$, which leads to the energy of the interface modes to be $E_i = \sqrt{1+u^2}$ which is identical for the symmetric and antisymmetric modes.
For a bridge interface, we have $h_1 = 1$ and $h_2=0$, which leads to the energy of the interface modes to be $E_i = \sigma+u$.
From this expression we can directly see that the interface mode for a bridge interface is the anti-symmetric mode.

\begin{figure}
    \centering
    \includegraphics[scale=0.85]{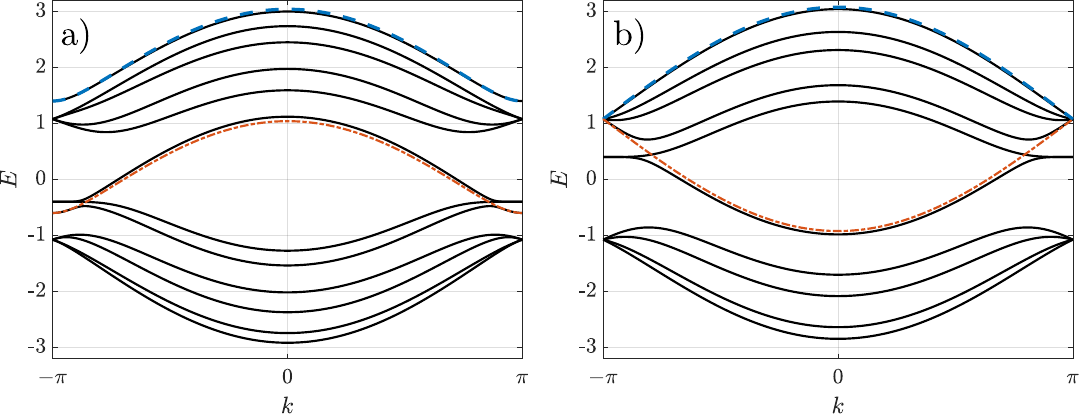}
    \caption{Comparison between the analytical expression for the energy of the interface states in infinite ribbons and numerical spectrum of a finite ribbon. The orange dotted-dashed curve is energy of the anti-symmetric interface state and the blue dashed curve is the one of the symmetric interface states. Panel (a) shows the comparison for a bridge interface with $N=6$. Panel (b) shows the comparison for a zigzag interface with $N=5$. In both cases, we have $u=0.4$.}
    \label{fig:app_infinite_vs_finite}
\end{figure}

To see that it is also the anti-symmetric mode in the case a zigzag interface, one has to compare the dispersion relation for all values of $k$ and not only in the fully dimerized limit.
A comparison between Eq.~\eqref{eq:energy_interface} and the spectrum of a finite chain is depicted in Fig.~\ref{fig:app_infinite_vs_finite}.

First of all, we can see that the approximation of an infinite chain reproduces closely the spectrum of interface states for a finite ribbon, even for relatively small values of $N$. Fig.~\ref{fig:app_infinite_vs_finite} shows the comparison for $N=5$ and $N=6$.
Secondly, we see that for both interfaces, it is the anti-symmetric interface modes which lies in the band gap.
This explain the observation of the interface modes presented in the main text.

\section{An example where valley change is unavoidable}\label{sec:Shift}

\begin{figure}[h]
    \centering
    \includegraphics[scale = 0.85]{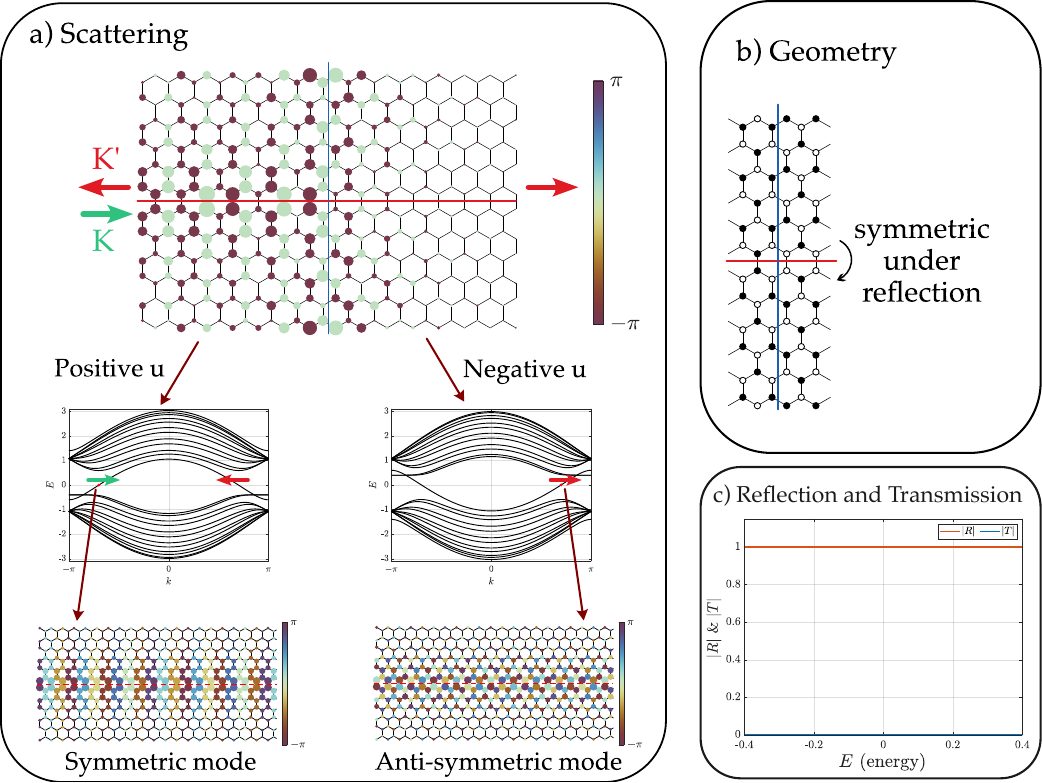}
    \caption{Scattering in a straight ribbon with flip of the onsite potential. Panel a) shows a particular scattering configuration for a ribbon with $N=12$ and $u=0.4$ at energy $E=0$, together with the dispersion relation on each side of the defect and the mode profiles at $E=0$. The arrows represent the direction of propagation of the modes and their color indicates their valley index. Panel b) shows the defect cell, the red and blue line corresponds to the interfaces where the onsite potential changes sign.
    Panel c) shows the reflection and transmission coefficients spectrum for the entire energy range where the valley Hall modes are present.}
    \label{fig:switch}
\end{figure}

We consider here a simple configuration where the valley index necessarily changes during the scattering process. The configuration consists in a straight ribbon with an interface (shown in red line in Fig.~\ref{fig:switch}) where the onsite potential $u>0$ is suddenly switched to $-u$. This abrupt change of the onsite potential happens at the blue interface in Fig.~\ref{fig:switch}.
\\
In Fig.~\ref{fig:switch}-a), we show the dispersion relation on both side of the defect, and we can directly see that the spectrum of the two ribbons are opposite. This implies in particular that if the right-moving valley Hall mode in the ribbon with $u>0$ has momentum $k$, then it will have momentum $-k$ in the ribbon with $u<0$, i.e. it will be located in the opposite valley.
In addition, right-moving and left-moving valley Hall modes are located, by construction in opposite valleys.
Hence, any scattering in such configuration will consist in an incoming mode in a specific valley, being either transmitted or reflected to modes with an opposite valley index. Therefore, a change in the valley index is unavoidable and there is a priori no way of distinguishing the preferred scattering channel between the transmitted and the reflected mode.
\\
Further insights can nonetheless be gained by inspecting the modes on each side of the defect. As shown in the supplemental material Sec.~\ref{App:RM}, interface modes will have symmetric or anti-symmetric profiles, depending on the type of interface and the sign of the onsite potential. This is illustrated in Fig.~\ref{fig:switch}-a) where the right-moving modes are shown on either side of the ribbon. We can see in particular that to the left of the defect, the valley Hall modes are symmetric across the red interface while they are anti-symmetric to the right of the defect.
Therefore, scattering from one side to the defect to the other will require not only to change the valley index but also to change the symmetry of the modes. Since the defect is itself mirror symmetric across the horizontal interface, see Fig.~\ref{fig:switch}-b), it cannot mix the symmetric and anti-symmetric components of the modes. Therefore, we can anticipate, from symmetry consideration, that a incoming mode will be fully reflected at the switching interface. This is indeed what is observed in Fig.~\ref{fig:switch}-a) and c).

\section{Trivial vs topological modes}\label{Sec:Trivial}
One of the main results of this study is the observation of a high transmission of interface modes across defects independently of the conservation of the valley index. 
This observation suggests that the topological nature of the modes involved does not determine the amount of backscattering. 
\begin{figure}[h]
    \centering
    \includegraphics[scale=0.85]{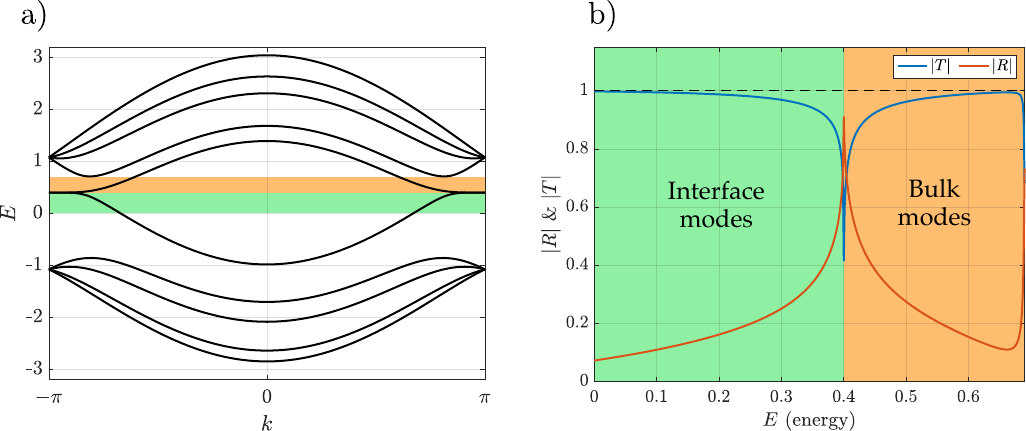}
    \caption{Comparison between interface and bulk modes scattering in a sharp bend at angle $2\pi/3$ in a ribbon with $N=5$ and $u=0.4$. Panel a) shows the dispersion relation of the ribbon, the green region shows part of the energy range where interface modes are present and the orange region shows the region where a single pair of bulk modes are propagating. Panel b) shows the scattering coefficients for both interface and bulk modes, we can see that the amount of backscattering is similar for the two types of modes.}
    \label{fig:trivial}
\end{figure}
\\
Here, we compare the scattering coefficients of a interface valley Hall mode with these of a trivial bulk mode of the ribbon, in the case of a sharp turn at angle $2\pi/3$ for a ribbon with $N=5$ and $u=0.4$. The dispersion relation for the ribbon and the scattering coefficients are shown in Fig.~\ref{fig:trivial}. Here we have chosen a relatively narrow ribbon to limit the number of propagating modes. This allows us to define an energy range where a single pair of propagating bulk modes is present, shown in orange. This energy range is comparable to the energy range shown in green where a single pair of propagating interface modes is present.
We can see in Fig.~\ref{fig:trivial}-b) that the scattering coefficients are similar for both interface and bulk modes, reinforcing our observation that the amount of backscattering is not characterised by topological properties.

\section{Extension to other interface types}\label{Sec:armchair}

In the main part of the text, we have constructed and used the transfer matrix formalism to quantitatively analyse the scattering of valley Hall modes through sharp bends in graphene ribbons with zigzag edges.
The construction of the propagating and evanescent modes via the transfer matrix presented in Sec.~\ref{sec:TM_ribbons} was specific to this particular ribbon configuration with zigzag edges.
We show here that it is possible to generalise the construction such that it can be applied to arbitrary ribbons. As an illustration, we then apply it to study the scattering of interface modes in graphene ribbons with armchair edges.

\subsection{Transfer matrix for arbitrary ribbons}

The construction of the transfer matrix in Sec.~\ref{sec:TM_ribbons} was done directly from the specific ribbon configuration and its decomposition in tilted supercells.
In particular, we used the fact the supercell only contained sites which were connected to adjacent supercells.

Here, we present a more abstract discussion allowing us to construct the transfer matrix of arbitrary ribbons with arbitrary supercell.
This generalisation comes at very little cost, since most of the necessary tools were introduced when discussing the connection through the bend in Sec.~\ref{sec:TM_bend}.

Similarly to Sec.~\ref{sec:TM_bend}, we introduce the vector $\Phi_n$ which contains the amplitudes on the sites of the $n$-th supercell of the ribbons. 
Since the ribbon is now arbitrary, there is no guarantee that the supercell will only be composed of sites directly connected to the adjacent supercells. Therefore, the vector $\Phi_n$ can be constructed from the vectors $\mathcal{A}_n$, $\mathcal{B}_n$ and $\mathcal{X}_n$, where
\begin{itemize}
    \item $\mathcal{A}_n$ contains the amplitudes on the sites which belong to the left edge of the supercell and are connected to the $n-1$ supercell.
    \item $\mathcal{B}_n$ contains the amplitudes on the sites which belong to the right edge of the supercell and are connected to the $n+1$ supercell.
    \item $\mathcal{X}_n$ contains the amplitudes on the internal sites which are neither connected to the $n-1$ nor $n+1$ supercell.
\end{itemize}
Together with these various sets, we define their associated projector operators as in Eq.~\eqref{eq:projectors}.
Therefore, $\Phi_n$ can be decomposed as
\begin{equation}
    \Phi_n = \vec{P}_\mathcal{A} \mathcal{A}_n + \vec{P}_\mathcal{B} \mathcal{B}_n + \vec{P}_\mathcal{X} \mathcal{X}_n.
\end{equation}

As for the bend cell, we write the eigenvalue equation as 
\begin{equation}
    \vec{P}_\mathcal{A}\cdot \vec{P}_\mathcal{B}^T \Phi_{n-1} + \bm{H}_0 \Phi_n + \vec{P}_\mathcal{B}\cdot \vec{P}_\mathcal{A}^T \Phi_{n+1} = E\Phi_n,
\end{equation}
where $\bm{H}_0$ is the Hamiltonian of the isolated supercell.
Following similar steps as in Sec.~\ref{sec:TM_bend} and defining the Green function of the supercell $\bm{\mathcal{G}} = \left( E - \bm{H}_0\right)^{-1}$, we arrive at equations analogous to Eqs.~\eqref{eq:connection1},~\eqref{eq:connection2} and~\eqref{eq:connection3}, namely
\begin{eqnarray}
    \mathcal{A}_n &=& \bm{\mathcal{G}}_{ab} \mathcal{A}_{n+1} + \bm{\mathcal{G}}_{aa} \mathcal{B}_{n-1}, \label{eq:connection1_bis}\\
    \mathcal{B}_n &=& \bm{\mathcal{G}}_{bb} \mathcal{A}_{n+1} + \bm{\mathcal{G}}_{ba} \mathcal{B}_{n-1}, \label{eq:connection2_bis}\\
    \mathcal{X}_n &=& \bm{\mathcal{G}}_{xb} \mathcal{A}_{n+1} + \bm{\mathcal{G}}_{xa} \mathcal{B}_{n-1}.\label{eq:connection3_bis}
\end{eqnarray}
In the above equations, $\bm{\mathcal{G}}_{aa}$, $\bm{\mathcal{G}}_{ab}$, etc... are the block components of $\bm{\mathcal{G}}$ and are obtained via the application of the appropriate projectors, as in Sec.~\ref{sec:TM_bend}.

We now connect with Sec.~\ref{sec:TM_ribbons} and introduce the vector $\Psi_n$ and the two matrices $\bm{M}_1$ and $\bm{M}_2$ defined as
\begin{equation}\label{eq:def_M1M2}
    \Psi_n = \binom{\mathcal{A}_n}{\mathcal{B}_{n-1}}, \quad 
    \bm{M}_1 = \begin{pmatrix}
        \bm{\mathcal{G}}_{ab} & 0 \\
        -\bm{\mathcal{G}}_{bb} & \bm{1}
    \end{pmatrix},\quad \text{and} \quad
    \bm{M}_2 = \begin{pmatrix}
        \bm{1} & -\bm{\mathcal{G}}_{aa} \\
        0 & \bm{\mathcal{G}}_{ba}
    \end{pmatrix}.
\end{equation}
For the new vector $\Psi_n$, the governing equation can be written in the form
\begin{equation}\label{GEN_rec}
    \bm{M}_1 \Psi_{n+1} = \bm{M}_2 \Psi_{n}.
\end{equation}
In order to define the transfer matrix of the ribbon, we have to invert the matrix $\bm{M}_1$. Due to its block form, $\bm{M}_1$ is invertible as long as $\bm{\mathcal{G}}_{ab}$ in invertible. 
We can see that there are two potential complications when constructing the transfer matrix : i) the matrix $(E - \bm{H}_0)$ may not be invertible and ii) the $\bm{\mathcal{G}}_{ab}$ block may not be invertible.
The first complication will arise if $E$ is in the spectrum of $\bm{H}_0$ which is a finite and discrete set of energies.
The second problem may occur for other values of the energy and has to be examined for the specific problem one is interested in.
Assuming that none of the above complication arises, which is numerically tested, Eq.\eqref{GEN_rec} can be simplified to
\begin{equation}
    \Psi_{n+1} = \bm{M} \Psi_n,
\end{equation}
where
\begin{equation}\label{eq:general_TM}
    \bm{M} = \begin{pmatrix}
        \bm{\mathcal{G}}_{ab}^{-1} & -\bm{\mathcal{G}}_{ab}^{-1}\cdot \bm{\mathcal{G}}_{aa} \\
        \bm{\mathcal{G}}_{bb}\cdot\bm{\mathcal{G}}_{ab}^{-1} & \bm{\mathcal{G}}_{ba} - \bm{\mathcal{G}}_{bb}\cdot\bm{\mathcal{G}}_{ab}^{-1}\cdot\bm{\mathcal{G}}_{aa}
    \end{pmatrix}.
\end{equation}
$\bm{M}$ is the transfer matrix of the graphene ribbon and its eigenvalues/eigenmodes contain all the information about the various modes present in the system.
From this matrix, one can construct all the propagating and evanescent modes in a similar way to the zigzag ribbon case of Sec.~\ref{sec:TM_ribbons}.

\subsubsection{Connection with graphene ribbons with zigzag edges}

As an illustration, we reconnect with the case of the graphene ribbon with zigzag edge of Sec.~\ref{sec:TM_ribbons} and construct its transfer matrix following the general construction presented above. For simplicity, we will consider graphene ribbon with zigzag edges and no onsite potential. With the notation of Sec.~\ref{sec:TM_ribbons}, this implies $\bm{U}_A = \bm{U}_B = 0$.

To construct the transfer matrix from Eq.~\eqref{eq:general_TM}, we first need the block components of the Green function $(E-\bm{H}_0)^{-1}$ and the matrices $\bm{M}_1$ and $\bm{M}_2$.
These can be obtained by putting Eqs.~\eqref{eq:AnBn1} and~\eqref{eq:AnBn2} in the form of Eqs.~\eqref{eq:connection1_bis} and~\eqref{eq:connection2_bis}. After some manipulation of Eqs.~\eqref{eq:AnBn1} and~\eqref{eq:AnBn2} we get
\begin{eqnarray}
\left[ E^2 - (\id + \bm{J}^T)(\id + \bm{J}) \right]\mathcal{A}_n &=& (\id + \bm{J})\mathcal{A}_{n+1} +  E \mathcal{B}_{n-1},\\
    \left[ E^2 - (\id + \bm{J}^T)(\id + \bm{J}) \right]\mathcal{B}_n &=& E\mathcal{A}_{n+1} + (\id + \bm{J}^T) \mathcal{B}_{n-1}.
\end{eqnarray}
We see that it is possible to identify the above equation with Eqs.~\eqref{eq:connection1_bis} and~\eqref{eq:connection2_bis} if 
\begin{equation}
    \text{det}\left[ E^2 - (\id + \bm{J}^T)(\id + \bm{J}) \right] \neq 0.
\end{equation}
The values of $E$ for which the above condition is not satisfied is precisely the spectrum of the isolated supercell of the graphene ribbon with zigzag edges.
Indeed, labelling the sites of the supercell as $(\mathcal{A}_n \ \mathcal{B}_n)^T$, the Hamiltonian of the supercell is given by
\begin{equation}
    \bm{H}_0 = \begin{pmatrix}
        \bm{0} & \bm{Q} \\
        \bm{Q}^T & \bm{0}
    \end{pmatrix}, \quad \text{with} \quad \bm{Q} = \id+\bm{J}^T.
\end{equation}
Therefore, the spectrum of $\bm{H}_0$ is given by the condition
\begin{eqnarray}
    \text{det}\left[E-\bm{H}_0\right] =  \text{det}\left[E^2-\bm{Q}\bm{Q}^T\right] = \text{det}\left[E^2-(\id+\bm{J}^T)(\id+\bm{J})\right] = 0.
\end{eqnarray}
Provided that $E$ is not an eigenvalue of $H_0$, we can identify the block components of the Green function 
\begin{eqnarray}
    \bm{\mathcal{G}}_{aa} &=& E\left[E^2-\bm{Q}\bm{Q}^T\right]^{-1},\\
    \bm{\mathcal{G}}_{ab} &=& \left[E^2-\bm{Q}\bm{Q}^T\right]^{-1}\cdot (\id+\bm{J}^T),\\
    \bm{\mathcal{G}}_{bb} &=& E\left[E^2-\bm{Q}\bm{Q}^T\right]^{-1},\\
    \bm{\mathcal{G}}_{ba} &=&  \left[E^2-\bm{Q}\bm{Q}^T\right]^{-1}\cdot (\id+\bm{J}).
\end{eqnarray}
From the above expressions, we can see that as long as $\bm{\mathcal{G}}_{ab}$ is well defined it will always be invertible and one can therefore define the transfer matrix. 
From the above expression, we can then reconstruct the matrices $\bm{M}_1$ and $\bm{M}_2$ in Eq.~\eqref{eq:def_M1M2}. 

\subsection{Application to ribbons with armchair edges}

We have now constructed the transfer matrix for general ribbons with arbitrary supercell. As an illustration we now apply this formalism to the case of graphene ribbons with armchair edges. 

\subsubsection{Graphene ribbons}

\begin{figure}
    \centering
    \includegraphics[scale=0.8]{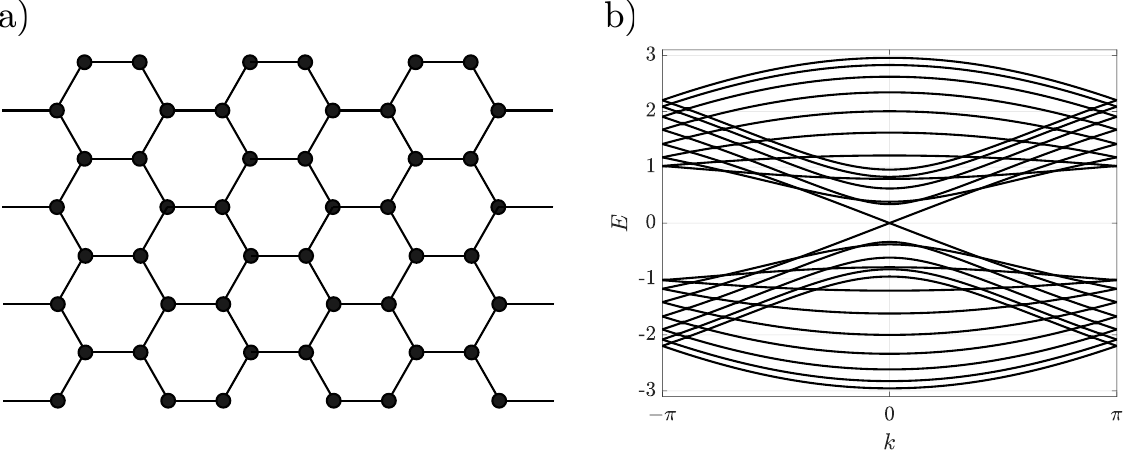}
    \caption{Graphene ribbon with armchair edges (Panel a) and its band structure (Panel b). The band structure is for a ribbon with width $N=14$ and is chosen such that the gap closes at $k=0$~\cite{wakabayashi2010electronic}.}
    \label{fig:armchair}
\end{figure}

As for the case of the zigzag ribbon, we begin by briefly reviewing the band properties of the armchair ribbons without the addition of onsite potential. The ribbon is depicted in Fig.~\ref{fig:armchair}.

The energy bands of the armchair ribbon without onsite potential can be expressed analytically in closed form\cite{wakabayashi2010electronic}. They are labelled by the transverse momenta $p_n$ which are determined from the width of the ribbon and take the form 
\begin{equation}
    p_n = \frac{n}{N+1} \pi,  \quad \text{with} \quad  n = 1,2,...,N.
\end{equation}
The energy bands are then given by
\begin{equation}\label{eq:disp_armchair}
    E_n(k) = \pm \sqrt{1 + 2 \cos(p_n)\cos\left(\frac{k}{2}\right) + 4 \cos(p_n)^2}.
\end{equation}

Since the transverse momentum $p_n$ does not depend on the longitudinal Bloch momentum $k$, we can obtain Eq.~\eqref{eq:disp_armchair} by slicing the 2-dimensional dispersion relation of the graphene lattice, along $k_x = 2p$ and setting $k = \sqrt{3} ky$ (according to the orientation and dimensions of the lattice as in Fig.~\ref{fig:graphene_nogap_summary}).
In particular, we see that the two Dirac points, $K$ and $K'$ are now superimposed at the Bloch wavenumber $k=0$ and the two valleys are not well separated in the Brillouin zone of the armchair ribbon.
It should also be noted that armchair ribbons have a Dirac point with no gap when $N=3l - 1$, with $l \in \mathbb{N}$ and a gap otherwise.
The dispersion relation for a ribbon with $N=14$ is shown in Fig.~\ref{fig:armchair}-b).

It is also known that the armchair edge is the only edge type for graphene ribbons which does not support edge state~\cite{delplace2011zak}.

\begin{figure}
    \centering
    \includegraphics[scale=0.8]{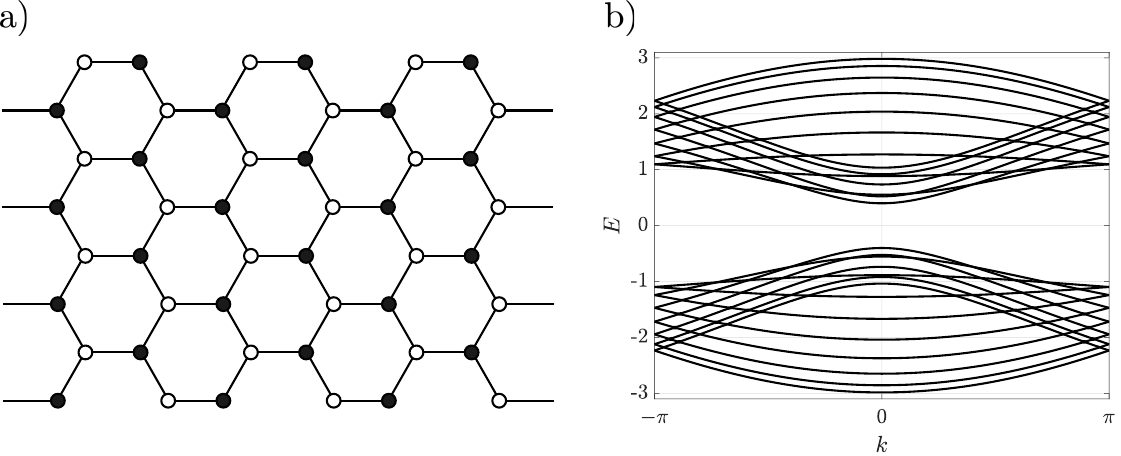}
    \caption{Graphene ribbon with armchair edges and onsite potential (Panel a) and its band structure (Panel b). The band structure is for a ribbon with width $N=14$ and onsite potential $u=0.4$.}
    \label{fig:armchair2}
\end{figure}

As in the case of the ribbon with zigzag edges, the addition of an onsite potential does not induce fundamental qualitative differences. In a case of a ribbon with no band gap, the addition of an onsite potential opens a gap at $k=0$. The armchair ribbon with onsite potential still does not support edge states.
The armchair ribbon with an onsite potential and its band structure are shown in Fig.~\ref{fig:armchair2}. 

\subsubsection{Ribbons with an interface}

We have recalled that graphene ribbons with armchair edges do not support edge states. However, as in the case of the zigzag edges, it is possible to construct armchair ribbons with an interface. Such ribbons will allow for the existence of localised modes at the interface.
It is possible to define scattering scenarios with the tools we constructed in the main part of the text to analyse the scattering of interface modes in armchair ribbons through different bends.

\begin{figure}
    \centering
    \includegraphics[scale=0.8]{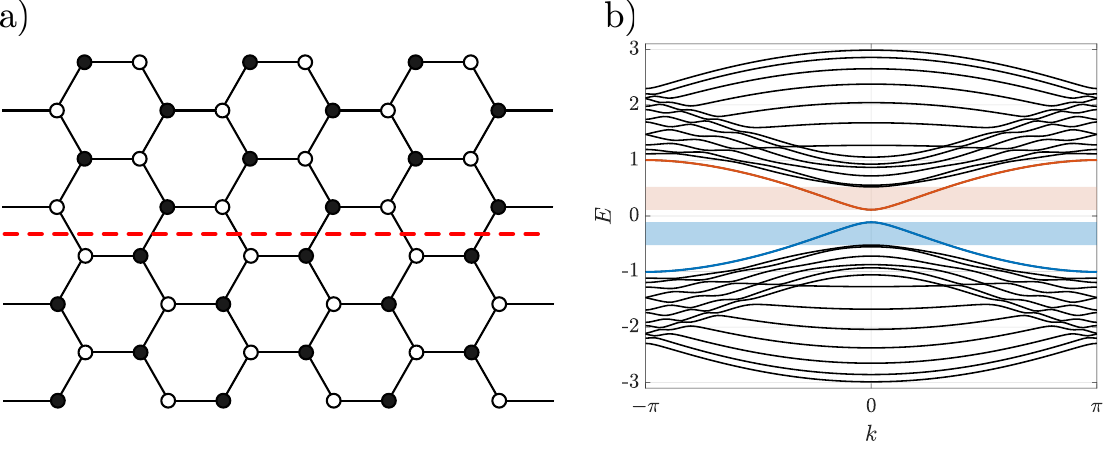}
    \caption{Armchair ribbon with an interface (Panel a) and its band structure (Panel b). The blue and ref curves represent the bands associated with modes localised at the interface of the armchair ribbon. The blue and red shaded regions refer to the energy range where only the interface modes are propagating. The parameters are chosen as $N=14$ and $u=0.4$.}
    \label{fig:armchair_interface}
\end{figure}

Fig.~\ref{fig:armchair_interface}-a) shows an armchair ribbon with an interface and its associated band structure is depicted in Fig.~\ref{fig:armchair_interface}-a) for a ribbon with $N=14$ and $u=0.4$.
The blue and red curves represent the bands associated with localised interface states and the shaded regions correspond to energy ranges where only the interface modes are propagating.
For such energies, we can apply the same formalism and definitions as in the case of the zigzag edge to define scattering coefficient in the case of interface modes of an armchair ribbon going through different bends.

\begin{figure}
    \centering
    \includegraphics[scale = 0.6]{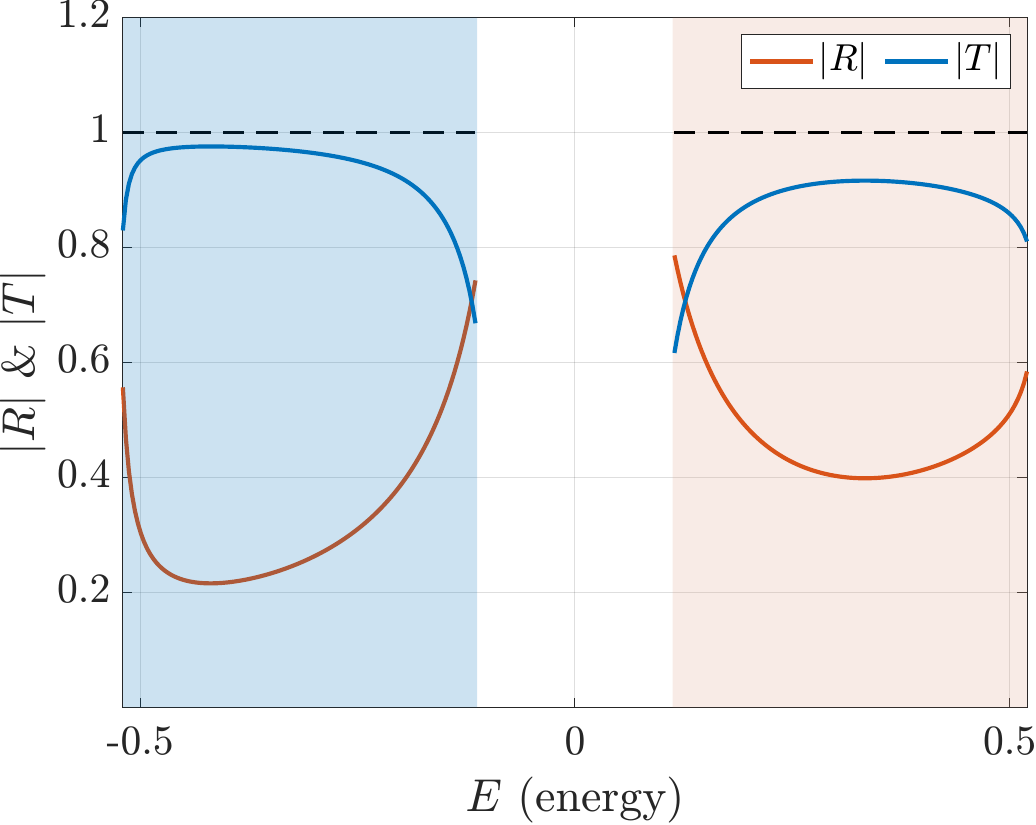}
    \caption{Reflection and transmission spectrum for the armchair ribbon with an interface with $N=14$ and $u=0.4$ accross a bend of angle $\pi/3$. The red (blue) curve represents the reflection (transmission) coefficient. The dashed black curve shows the combination $|R|^2 + |T|^2$. The shaded colored regions correspond to the energy ranges depicted in Fig.~\ref{fig:armchair_interface} b).}
    \label{fig:bad_armchair}
\end{figure}

Fig.~\ref{fig:bad_armchair} shows the reflection and transmission coefficients of interface modes in armchair ribbons with an interface scattering through a bend with angle $\pi/3$. We can see that the spectrum is gapped around $E=0$ and the shaded regions correspond to the energy ranges shown in Fig.~\ref{fig:armchair_interface}. We can see that the spectrum in the two separate energy ranges are qualitatively similar to the one for a ribbon with zigzag edges through a similar bend. Quantitatively, the amount of reflection appears larger for the armchair ribbon and this set of parameters.

The interpretation and link to the valley Hall effect of this particular configuration is however distinct from the zigzag edges case. This case deserves further investigation which is left for future studies.

\end{document}